\documentclass[ reprint,
amsmath,amssymb,aps,pra]{revtex4-1}

\usepackage{graphicx}
\usepackage{dcolumn}
\usepackage{bm}

\begin{document}

\preprint{APS/123-QED}

\title{Classification of propagation-invariant space-time wave packets in free space:\\
Theory and experiments}

\author{Murat Yessenov}
\author{Basanta Bhaduri}
\author{H. Esat Kondakci}
\author{Ayman F. Abouraddy}
\email{raddy@creol.ucf.edu}
\affiliation{CREOL, The College of Optics \& Photonics, University of Central Florida, Orlando, FL 32816, USA}%

\date{\today}

\begin{abstract}
Introducing correlations between the spatial and temporal degrees of freedom of a pulsed optical beam (or wave packet) can profoundly alter its propagation in free space. Indeed, appropriate spatio-temporal spectral correlations can render the wave packet propagation-invariant: the spatial \textit{and} temporal profiles remain unchanged along the propagation axis. The spatio-temporal spectral locus of any such wave packet lies at the intersection of the light-cone with tilted spectral hyperplanes. We investigate $(2\!+\!1)$D propagation-invariant `space-time' light sheets, and identify 10 classes categorized according to the magnitude and sign of their group velocity and the nature of their spatial spectrum -- whether the low spatial frequencies are physically allowed or forbidden according to their compatibility with causal excitation and propagation. We experimentally synthesize and characterize all 10 classes using an experimental strategy capable of synthesizing space-time wave packets that incorporate arbitrary spatio-temporal spectral correlations. 
\end{abstract}

\pacs{Valid PACS appear here}
\maketitle

\section{Introduction}

Optical diffraction sets universal performance limits on microscopy, lithography, and imaging, among myriad other areas. This fundamental limitation has motivated a long-standing effort for developing strategies to combat diffractive spreading \cite{Airy41PM,Rayleigh72MNRAS,Steward26PTRCL,Stratton41Book,Steel53RO,Welford60JOSA,Sheppard77Optik}, culminating in so-called `diffraction-free' beams \cite{Durnin87PRL,Bandres04OL,Rodriguez10JOSAA,Siviloglou07PRL,Levy16PO}, of which Airy beams are the only scalar 1D diffraction-free optical \textit{sheet} \cite{Berry79AMP,Siviloglou07OL}. In considering \textit{pulsed} beams (or wave packets), propagation invariance in free space has been predicted for specific wave packets, including Brittingham's focus-wave mode (FWM) \cite{Brittingham83JAP}, Mackinnon's wave packet \cite{Mackinnon78FP}, X-waves \cite{Lu92IEEEa,Lu92IEEEb,Saari97PRL}, among many others \cite{Besieris89JMP,Wunsche89JOSAA,Donnelly93PRSLA,Besieris98PIERS,Salo00PRE,Sheppard02JOSAA} (see \cite{Turunen10PO,FigueroaBook14} for reviews).

Underlying the propagation invariance of these wave packets is a fundamental principle: the \textit{spatial} frequencies involved in the construction of the beam profile must be correlated with the \textit{temporal} frequencies (wavelengths) underlying the pulse linewidth \cite{Donnelly93PRSLA,Longhi04OE,Saari04PRE,Kondakci16OE,Kondakci17NP}. Consequently, the spectrum is confined to a \textit{reduced-dimensionality} space with respect to traditional wave packets \cite{Kondakci17NP}, and we refer to them as `space-time' (ST) wave packets \cite{Kondakci16OE,Parker16OE}. We do not consider here scenarios in which chromatic dispersion \cite{Sonajalg97OL,Porras01OL,Orlov02OL,Lu03JOSAA,Christodoulides04OE,Longhi04OL,Dallaire09OE,Jedrkiewicz13OE} or optical nonlinearities \cite{DelRe15NP} are required for propagation invariance.

Substantial efforts have been directed to the synthesis of ST wave packets exploiting the traditional approaches for producing Bessel beams (e.g., annular apertures \cite{Saari97PRL} and axicons \cite{Bonaretti09OE,Bowlan09OL,Alexeev02PRL}, via nonlinear interactions \cite{DiTrapani03PRL,Faccio06PRL,Faccio07OE}, or spatio-temporal filtering \cite{Dallaire09OE,Jedrkiewicz13OE}. Recently, we introduced a spatio-temporal \textit{synthesis} strategy capable of preparing any $(2\!+\!1)$D ST wave packet in the form of pulsed sheets through the use of a spatial light modulator (SLM) to perform joint spatio-temporal spectral shaping of a femtosecond pulsed plane wave. By encoding a prescribed one-to-one correspondence between the spatial frequencies $|k_{x}|$ and temporal frequencies $\omega$ in the spatio-temporal spectrum, where $x$ is the transverse coordinate \cite{Kondakci17NP,Kondakci18unpub}, a linear one-to-one correspondence is established between $\omega$ and $k_{z}$, where $z$ is the axial coordinate (the $y$-dependence is dropped), which guarantees rigid translation of the wave packet envelope in free space. The precision and versatility of this approach has been confirmed by the quality of fit between theoretical predictions and measurements of ST wave packets with tailorable beam profiles \cite{Kondakci17NP}, Airy ST wave packets that accelerate in space-time \cite{Kondakci18PRL}, broadband ST wave packets produced using refractive phase plates \cite{Kondakci18OE}, self-healing after opaque obstructions \cite{Kondakci18OL}, and extended travel distances \cite{Bhaduri18OE}. Additionally, recent theoretical studies have also examined the properties of such unique optical fields \cite{Wong17ACSP1,Wong17ACSP2,Porras17OL,SaintMarie17Optica,Efremidis17OL,PorrasPRA18}.

\begin{figure*}[t!]
  \begin{center}
  \includegraphics[width=14.5cm]{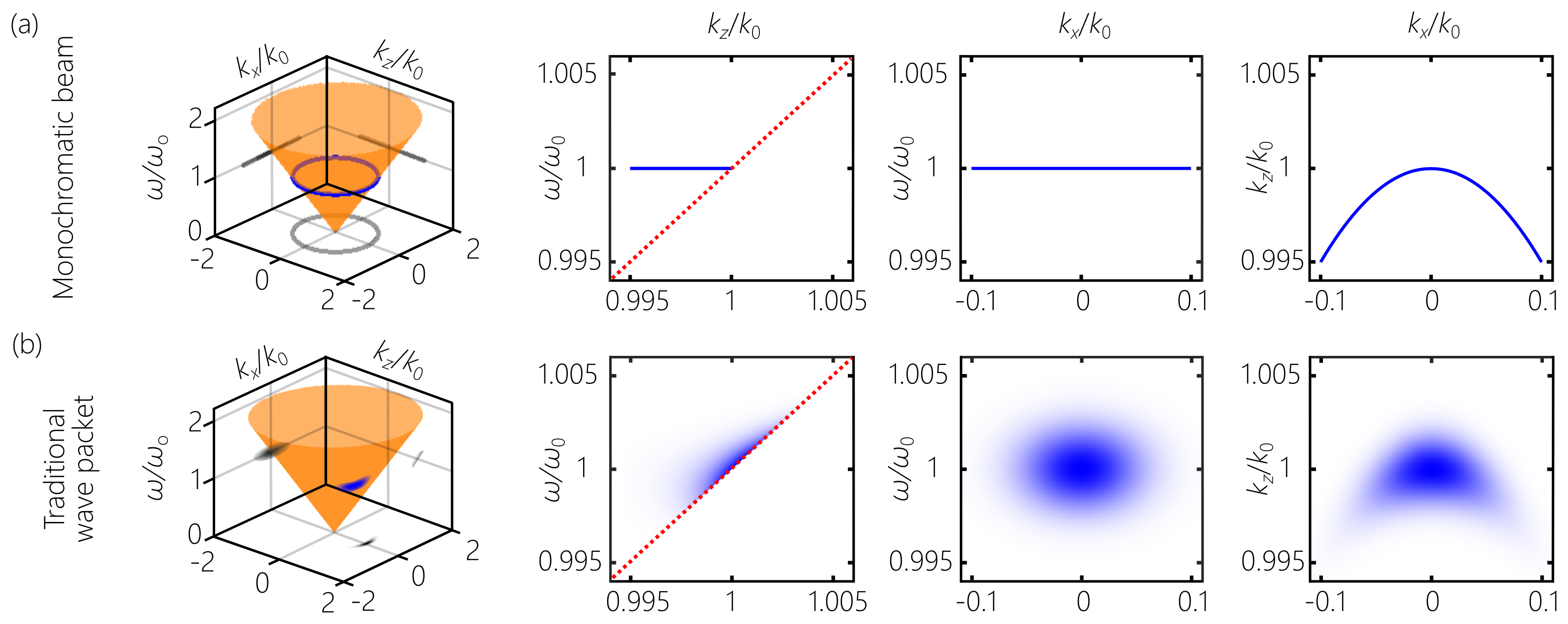}
  \end{center}
  \caption{Spatio-temporal spectra of traditional optical fields and their projections onto the $(k_{z},\tfrac{\omega}{c})$, $(k_{x},\tfrac{\omega}{c})$, and $(k_{x},k_{z})$ planes. (a) The spatial spectra of monochromatic beams lie at the intersection of the light-cone with the iso-frequency plane $\omega\!=\!\omega_{\mathrm{o}}$, where $k_{\mathrm{o}}\!=\!\tfrac{\omega_{\mathrm{o}}}{c}$. (b) The spatio-temporal spectra of traditional pulsed beams occupy 2D domains on the light-cone. We show a separable wave packet having a Gaussian beam profile and pulse linewidth. The red dotted line is the light-line $\tfrac{\omega}{c}\!=\!k_{z}$.}
  \label{Fig:Projections1-2}
\end{figure*}

In this paper we present a systematic classification of $(2\!+\!1)$D propagation-invariant ST wave packets and experimentally synthesize and characterize a representative from each of these classes. We identify 10 unique classes of ST wave packets indexed with respect to three criteria: (1) the group velocity -- subluminal, luminal, or superluminal; (2) the direction of the group velocity -- forward or backward with respect to the source; and (3) whether low-spatial frequencies are allowed or forbidden, which we refer to as `baseband' and `sideband' ST wave packets, respectively. Critically, we find that baseband ST wave packets offer more versatility compared to sideband ST wave packets (including X-waves) with regards to the tunability of their group velocity, which can be varied continuously to arbitrary values. Indeed, we demonstrate here the synthesis of baseband ST wave packets whose group velocities span the whole range of positive and negative subluminal, luminal, and superluminal values in free space. Previously published results deviate only slightly from the speed of light $c$: $1.00022c$ \cite{Bonaretti09OE}, $1.00012c$ \cite{Bowlan09OL}, $1.00015c$ \cite{Kuntz09PRA}, and $0.999c$ \cite{Lohmus12OL,Piksarv12OE} (there have been no observations to date of negative group velocities for such wave packets). On the other hand, controlling the group velocity of sideband ST wave packets is challenging due to the large spatial frequencies to be included in the spectrum. This possibility of synthesizing an arbitrary ST wave packet motivates providing the complete classification presented here \cite{Donnelly93PRSLA,Saari04PRE,Turunen10PO,PorrasPRA18}. 

We emphasize here the novel aspects of this work. First, we show that a single optical arrangement can synthesize \textit{any} class of ST wave packet, in contradistinction to previous efforts where distinct setups have been constructed for each class. Second, several of these classes are synthesized here for the first time; e.g., negative-group-velocity wave packets. Third, our linear, energy-efficient all-phase modulation scheme results in excellent quantitative agreement with theoretical predictions across the whole range of physically realizable parameters. Moreover, the simple geometric picture associated with our classification immediately reveals the physical limits set by causal excitation on realizable ST wave packets \cite{Heyman87JOSAA,Heyman89IEEE,Heyman93Chapter,Shaarawi95JMP}, and unveils, unexpectedly, a class of ST wave packets having a spectral range admitting of a two-to-one correspondence between $|k_{x}|$ and $\omega$.

The paper is organized as follows. First, we describe the origin of propagation invariance of an optical wave packet in the correlation between its spatial and temporal degrees of freedom. Next we present the criteria by which we classify ST wave packets before providing a detailed analysis of the 10 identified classes. We then describe our experimental methodology for synthesizing and analyzing ST wave packets, and present realizations of all 10 classes. We end the paper with a discussion of the opportunities for future development. 

\section{Spatio-temporal correlations as the basis for propagation-invariant space-time wave packets}\label{Sec:Conditions}

\begin{figure*}[t!]
  \begin{center}
  \includegraphics[width=16.5cm]{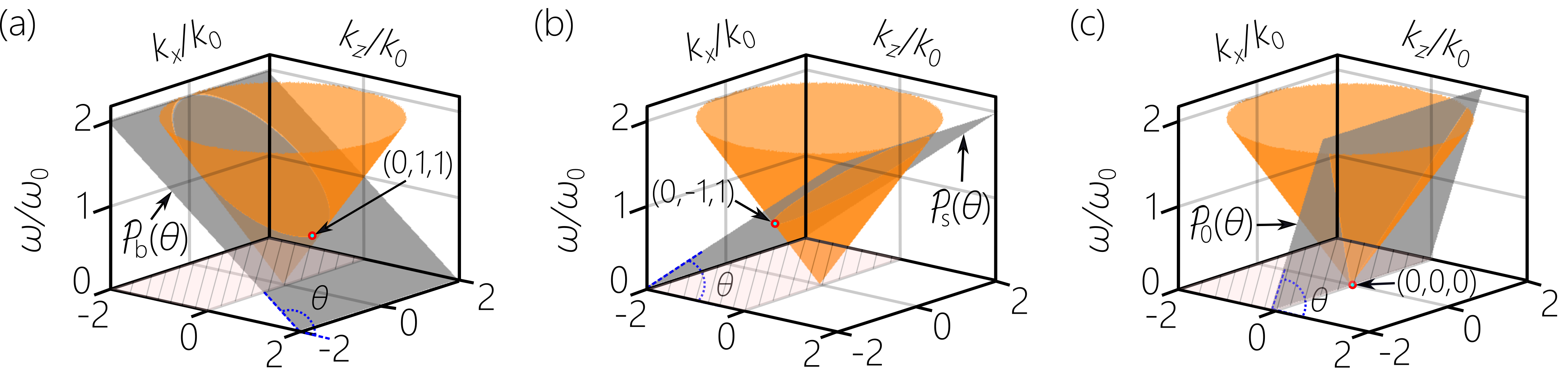}
  \end{center}
  \caption{Concept of ST wave packets in the spatio-temporal spectral domain. (a) \textit{Baseband} ST wave packets. The spatio-temporal spectra lie at the intersection of the light-cone with the spectral hyperplane $\mathcal{P}_{\mathrm{b}}(\theta)$ that passes through the point $(k_{x},k_{z},\tfrac{\omega}{c})\!=\!(0,k_{\mathrm{o}},k_{\mathrm{o}})$. (b) \textit{Sideband} ST wave packets. The spatio-temporal spectra lie at the intersection of the light-cone with the spectral hyperplane $\mathcal{P}_{\mathrm{s}}(\theta)$ that passes through the point $(k_{x},k_{z},\tfrac{\omega}{c})\!=\!(0,-k_{\mathrm{o}},k_{\mathrm{o}})$. (c) X-waves. The spatio-temporal spectra lie at the intersection of the light-cone with the spectral hyperplane $\mathcal{P}_{\mathrm{o}}(\theta)$ that passes through the origin $(k_{x},k_{z},\tfrac{\omega}{c})\!=\!(0,0,0)$. All hyperplanes are parallel to the $k_{x}$-axis, $\theta$ is the spectral tilt angle with the $(k_{x},k_{z})$-plane, and $k_{\mathrm{o}}$ is a fixed wave number. The shaded area ($k_{z}\!<\!0$) in each panel is excluded to guarantee causal generation of the ST wave packets.}
  \label{Fig:Concept}
\end{figure*}

In this Section we present a formalism for describing the spatio-temporal spectral loci of traditional and ST wave packets, emphasizing the difference of the dimensionality of the spectral representation on the light-cone. 

\subsection{Traditional optical wave packets}

We consider finite-bandwidth \textit{pulsed} fields $E(x,z,t)\!=\!\psi(x,z,t)e^{i(k_{\mathrm{o}}z-\omega_{\mathrm{o}}t)}$ that satisfy the homogeneous, linear, scalar wave equation $\{\partial^{2}_{x}\!+\!\partial^{2}_{z}\!-\!\tfrac{1}{c^{2}}\partial^{2}_{t}\}E\!=\!0$; here $k_{\mathrm{o}}\!=\!\omega_{\mathrm{o}}/c$ is a fixed wave number, the phase velocity is $v_{\mathrm{ph}}\!=\!c$, and we take the field distribution to be uniform along $y$. The envelope can be expressed in a plane-wave expansion,
\begin{equation}\label{Eq:GeneralPulsedBeamExpansion}
\psi(x,z,t)=
\!\int\!\!\!\int\!dk_{x}d\Omega\,\widetilde{\psi}(k_{x},\Omega)e^{i\{k_{x}x+(k_{z}-k_{\mathrm{o}})z-\Omega t\}},
\end{equation}
where the spatio-temporal spectrum $\widetilde{\psi}(k_{x},\Omega)$ is the 2D Fourier transform of $\psi(x,0,t)$ and $\Omega\!=\!\omega-\omega_{\mathrm{o}}$. The free-space dispersion relationship $k_{x}^{2}\!+\!k_{z}^{2}\!=\!(\tfrac{\omega}{c})^{2}$ implied in Eq.~\ref{Eq:GeneralPulsedBeamExpansion} corresponds geometrically to the surface of the `light-cone', and any monochromatic plane-wave $e^{i\{k_{x}x+k_{z}z-\omega t\}}$ is represented by a point $(k_{x},k_{z},\tfrac{\omega}{c})$ on its surface. For example, the spatial spectra of \textit{monochromatic beams} lie on the circle at the intersection of the light-cone with the iso-frequency plane $\omega\!=\!\omega_{\mathrm{o}}$ [Fig.~\ref{Fig:Projections1-2}(a)], whereas \textit{pulsed beams} occupy 2D domains on the surface of the light-cone [Fig.~\ref{Fig:Projections1-2}(b)]; see \cite{Kondakci16OE,Kondakci17NP}. Typical laser pulses are separable with respect to $k_{x}$ and $\omega$; however, exceptions in the case of high-energy pulses exist due to nonlinearities and aberrations \cite{Pariente16NP}. 

\subsection{The spectral locus of ST wave packets}

\textit{Rigid} translation (diffraction-free \textit{and} dispersion-free) of the ST wave-packet envelope necessitates a reduced-dimensionality form of the field with respect to that in Eq.~\ref{Eq:GeneralPulsedBeamExpansion}. Specifically, the spectral locus of such a ST wave packet must lie at the intersection of the light-cone with a tilted spectral hyperplane that is parallel to the $k_{x}$ -axis. The three planes of interest are shown in Fig.~\ref{Fig:Concept}: $\mathcal{P}_{\mathrm{b}}(\theta)$, $\mathcal{P}_{\mathrm{s}}(\theta)$, or $\mathcal{P}_{\mathrm{o}}(\theta)$, and are given by:
\begin{eqnarray}
\mathcal{P}_{\mathrm{b}}(\theta):\,\,\,\omega-\omega_{\mathrm{o}}&=&(k_{z}-k_{\mathrm{o}})\,c\tan{\theta},\\
\mathcal{P}_{\mathrm{s}}(\theta):\,\,\,\omega-\omega_{\mathrm{o}}&=&(k_{z}+k_{\mathrm{o}})\,c\tan{\theta},\\
\mathcal{P}_{\mathrm{o}}(\theta):\,\;\;\;\;\;\;\;\;\omega&=&k_{z}\,c\tan{\theta},
\end{eqnarray}
The projections of all three planes onto the $(k_{z},\tfrac{\omega}{c})$-plane is a straight line whose slope is $\tan{\theta}\!=\!v_{\mathrm{g}}/c$, where the group velocity $v_{\mathrm{g}}$ is determined solely by $\theta$, which is the angle with respect to the $k_{z}$-axis we refer to henceforth as the \textit{spectral tilt angle}. Consequently, the spatio-temporal spectrum of such a wave packet does \textit{not} occupy a 2D domain on the surface of the light-cone, such as that in Fig.~\ref{Fig:Projections1-2}(b), but lies instead along a 1D curve.

\textit{(1) Baseband ST wave packets:} The plane $\mathcal{P}_{\mathrm{b}}(\theta)$ passes through the point $(k_{x},k_{z},\tfrac{\omega}{c})\!=\!(0,k_{\mathrm{o}},k_{\mathrm{o}})$, and the subscript `b' indicates that such ST wave packets are `baseband' [Fig.~\ref{Fig:Concept}(a)]; i.e., low spatial frequencies down to $k_{x}\!=\!0$ are permissible. The field takes the form $E(x,z,t)\!=\!e^{i(k_{\mathrm{o}}z-\omega_{\mathrm{o}}t)}\psi(x,z-v_{\mathrm{g}}t)$, $v_{\mathrm{ph}}\!=\!c$, and the envelope can be expanded as
\begin{equation}\label{Eq:STExpansion}
\psi(x,z,t)\!=\!\int\!dk_{x}\,\tilde{\psi}(k_{x})\,e^{ik_{x}x}e^{i(k_{z}-k_{\mathrm{o}})(z-v_{\mathrm{g}}t)}.
\end{equation}

\begin{figure}[t!]
  \begin{center}
 \includegraphics[width=8.6cm]{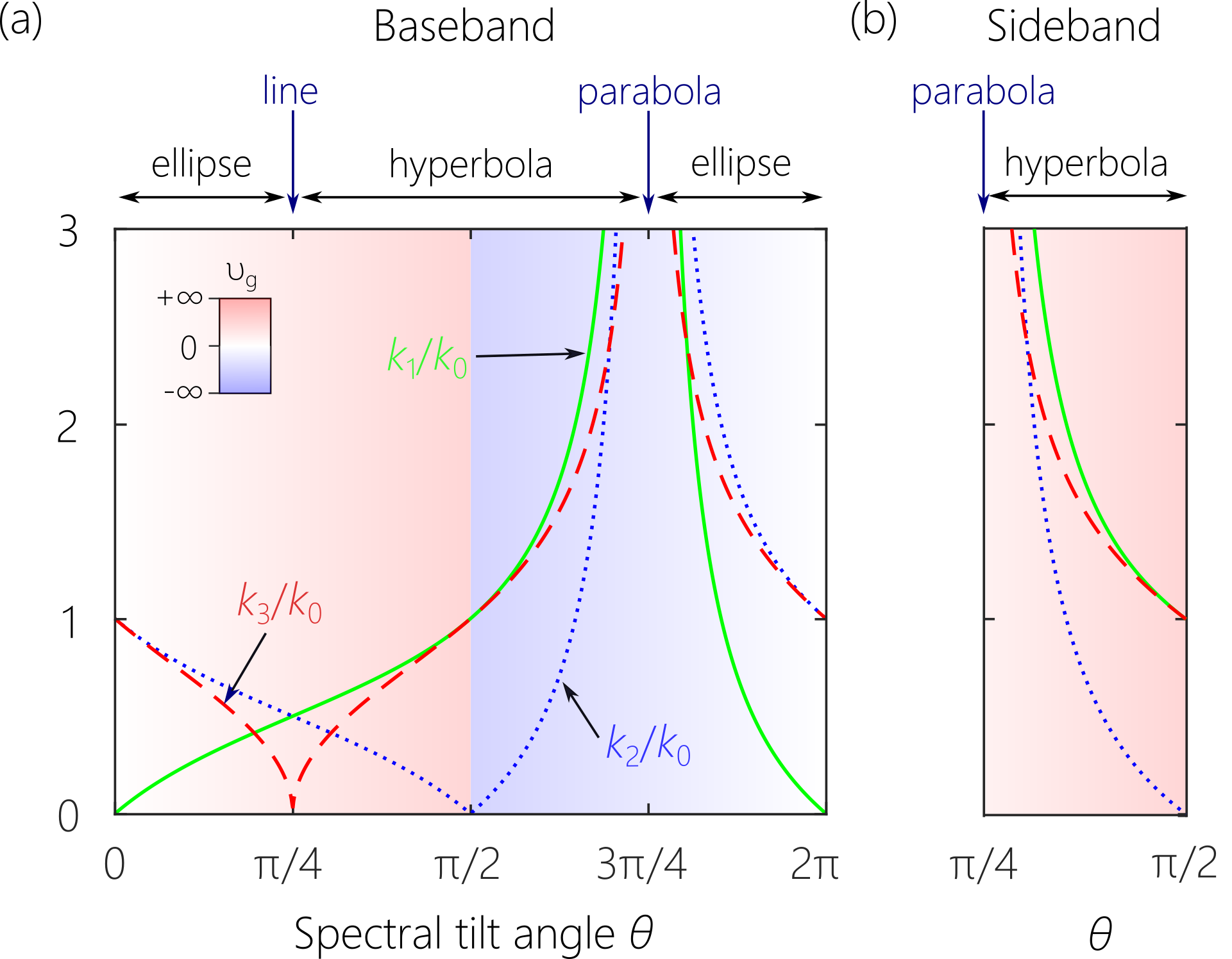}
  \end{center}
  \caption{(a) Plots of $k_{1}$, $k_{2}$, and $k_{3}$ (Eq.~\ref{Eq:k1k2k3}), and (b) $k_{1}'$, $k_{2}'$, and $k_{3}'$ (Eq.~\ref{Eq:k1k2k3dash}) with $\theta$. The background color indicates the corresponding value of $v_{\mathrm{g}}$ and the conic sections associated with the spatio-temporal spectra are indicated at the top.}
  \label{Fig:k1k2k3}
\end{figure}

The intersection of $\mathcal{P}_{\mathrm{b}}(\theta)$ with the light-cone is a conic section whose projections onto the $(k_{x},\tfrac{\omega}{c})$ and $(k_{x},k_{z})$ planes are
\begin{equation}\label{Eq:OmegaKxPlane}
\frac{1}{k_{1}^{2}}\left(\frac{\omega}{c}\pm k_{2}\right)^{2}\pm\frac{k_{x}^{2}}{k_{3}^{2}}=1,\,\,\,\frac{1}{k_{2}^{2}}\left(k_{z}\pm k_{1}\right)^{2}\pm\frac{k_{x}^{2}}{k_{3}^{2}}=1,
\end{equation}
respectively, where the signs are determined by $\theta$ as detailed below, and $k_{1}$, $k_{2}$, and $k_{3}$ are positive constants,
\begin{equation}\label{Eq:k1k2k3}
\frac{k_{1}}{k_{\mathrm{o}}}\!=\!\left|\frac{\tan{\theta}}{1\!+\!\tan{\theta}}\right|,\,\,\,\frac{k_{2}}{k_{\mathrm{o}}}\!=\!\frac{1}{\left|1\!+\!\tan{\theta}\right|},\,\,\,\frac{k_{3}}{k_{\mathrm{o}}}\!=\!\sqrt{\left|\frac{1\!-\!\tan{\theta}}{1\!+\!\tan{\theta}}\right|},
\end{equation}
plotted in Fig.~\ref{Fig:k1k2k3}(a). These relationships hold in the range $0\!<\!\theta\!<\!\pi$, and degenerate into those for a circle at $\theta\!=\!0$, a line at $\theta\!=\!\tfrac{\pi}{4}$, and a parabola at $\theta\!=\!\tfrac{3\pi}{4}$.

\textit{(2) Sideband ST wave packets:} The plane $\mathcal{P}_{\mathrm{s}}(\theta)$ passes through the point $(k_{x},k_{z},\tfrac{\omega}{c})\!=\!(0,-k_{\mathrm{o}},k_{\mathrm{o}})$, and the subscript `s' indicates that such ST wave packets are `sideband' [Fig.~\ref{Fig:Concept}(b)]; i.e., low spatial frequencies $k_{x}$ below a certain cutoff are incompatible with causal excitation and are thus forbidden (see below). The field takes the form $E(x,z,t)\!=\!e^{-i(k_{\mathrm{o}}z+\omega_{\mathrm{o}}t)}\psi(x,z-v_{\mathrm{g}}t)$, $v_{\mathrm{ph}}\!=\!-c$, and the plane-wave expansion is
\begin{equation}\label{Eq:STExpansionNegativePhaseVel}
\psi(x,z,t)\!=\!\int\!\!dk_{x}\,\widetilde{\psi}(k_{x})e^{ik_{x}x}e^{i(k_{z}+k_{\mathrm{o}})(z-v_{\mathrm{g}}t)},
\end{equation}

Planes $\mathcal{P}_{\mathrm{s}}(\theta)$ intersect with the light-cone in a parabola or a hyperbola having the form of Eq.~\ref{Eq:OmegaKxPlane}, but $k_{1}$, $k_{2}$ and $k_{3}$ are replaced with $k_{1}'$, $k_{2}'$, and $k_{3}'$, respectively, where
\begin{equation}\label{Eq:k1k2k3dash}
\frac{k_{1}'}{k_{\mathrm{o}}}\!=\!\left|\frac{\tan{\theta}}{1\!-\!\tan{\theta}}\right|,\,\,\,\frac{k_{2}'}{k_{\mathrm{o}}}\!=\!\frac{1}{\left|1\!-\!\tan{\theta}\right|},\,\,\,\frac{k_{3}'}{k_{\mathrm{o}}}\!=\!\sqrt{\left|\frac{1\!+\!\tan{\theta}}{1\!-\!\tan{\theta}}\right|},
\end{equation}
plotted in Fig.~\ref{Fig:k1k2k3}(b). Here the $\theta$ is restricted to $\tfrac{\pi}{4}\!<\!\theta\!<\!\tfrac{\pi}{2}$ as we show below.

\textit{(3) X-waves:} The plane $\mathcal{P}_{\mathrm{o}}(\theta)$ passes through the origin and intersects with the light-cone in a pair of lines $\omega/c\!=\!k_{z}\tan{\theta}$ [Fig.~\ref{Fig:Concept}(c)], leading to a field expansion:
\begin{equation}
E(x,z,t)=\int\!dk_{x}\tilde{\psi}(k_{x})e^{i\{k_{x}x+k_{z}(z-v_{\mathrm{g}}t)\}}.
\end{equation}
In principle the spatial spectrum can extend to include $k_{x}\!=\!0$, and thus corresponds to a \textit{baseband} ST wave packet, but this requires extending the temporal spectrum to $\omega\!=\!0$. In practice, this scenario therefore corresponds to \textit{sideband} ST wave packets. We show below that $\theta$ is restricted to the range $\tfrac{\pi}{4}\!<\!\theta\!<\!\tfrac{\pi}{2}$. Such ST wave packets are 1D analogs of X-waves \cite{Lu92IEEEa,Lu92IEEEb,Saari97PRL}.

\section{Classification of space-time wave packets}\label{Sec:AllBeam}

\begin{table*}[htb]\label{Table:Classification}
  \centering
  \caption{Classification of (2+1)D ST wave packets in free space.}
  \begin{tabular}
  {p{0.4in}p{0.7in}p{1.1in}p{0.5in}p{0.4in}p{0.7in}p{1in}p{1.8in}} \hline\hline
    \textbf{Class} & \textbf{$\theta$} & \textbf{Group velocity $v_{\mathrm{g}}$} & & \textbf{Sign} & \textbf{base/side} & \textbf{Conic section} & \textbf{Name [references]} \\ \hline\hline

\vspace{0.05cm}&&&&&\\
(1) & $0\!<\!\theta\!<\!\frac{\pi}{4}$ & Subluminal & $v_{\mathrm{g}}\!<\!c$ & +ve & baseband & ellipse & Mackinnon wave packet \cite{Mackinnon78FP,Wong17ACSP2} \\
&&&&&\\\hline
&&&&&\\
(2) & $\theta\!=\!\frac{\pi}{4}$ & Luminal & $v_{\mathrm{g}}\!=\!c$ & +ve & baseband & line & Pulsed plane wave\\
&&&&&\\\hline

&&&&&\\
(3) & $\frac{\pi}{4}\!<\!\theta\!<\!\frac{\pi}{2}$ & Superluminal & $v_{\mathrm{g}}\!>\!c$ & +ve & baseband & hyperbola & \cite{Valtna07OC,Wong17ACSP2} \\
&&&&&\\\hline

&&&&&\\
(4) &$\theta\!=\!\frac{\pi}{2}$ & Superluminal & $v_{\mathrm{g}}\!=\!\infty$ & --- & baseband & iso-$k_{z}$ hyperbola & \cite{Longhi04OE,Kondakci16OE,Parker16OE,Porras17OL,PorrasPRA18,Wong17ACSP1}\\
&&&&&\\\hline

&&&&&\\
(5) & $\frac{\pi}{2}\!<\!\theta\!<\!\frac{3\pi}{4}$ & Superluminal & $|v_{\mathrm{g}}|\!>\!c$ & -ve & baseband & hyperbola & ---\\
&&&&&\cite{Wong17ACSP1}\\\hline

&&&&&\\
(6) & $\theta\!=\!\frac{3\pi}{4}$ & Luminal & $|v_{\mathrm{g}}|\!=\!c$ & -ve & baseband & parabola & ---\\
&&&&&\\\hline

&&&&&\\
(7) & $\frac{3\pi}{4}\!<\!\theta\!<\!\pi$ & Subluminal & $|v_{\mathrm{g}}|\!<\!c$ & -ve & baseband & ellipse & \cite{Wong17ACSP2}\\
&&&&&\\\hline

&&&&&\\
(8) & $\theta\!=\!\frac{\pi}{4}$ & Luminal & $v_{\mathrm{g}}\!=\!c$ & +ve & sideband & parabola & Brittingham FWM \cite{Brittingham83JAP,Reivelt00JOSAA,Reivelt02PRE}\\
&&&&&\\\hline

&&&&&\\
(9) & $\frac{\pi}{4}\!<\!\theta\!<\!\frac{\pi}{2}$ & Superluminal & $v_{\mathrm{g}}\!>\!c$ & +ve & sideband & hyperbola & ---\\
&&&&&\\\hline

&&&&&\\
(10) & $\frac{\pi}{4}\!<\!\theta\!<\!\frac{\pi}{2}$ & Superluminal & $v_{\mathrm{g}}\!>\!c$ & +ve & sideband & lines & X-waves \cite{Lu92IEEEa,Lu92IEEEb,Saari97PRL}\\
&&&&&\\\hline\hline

\end{tabular}
\end{table*}

\subsection{Criteria for classifying ST wave packets}

\begin{figure*}[t!]
  \begin{center}
  \includegraphics[width=14.6cm]{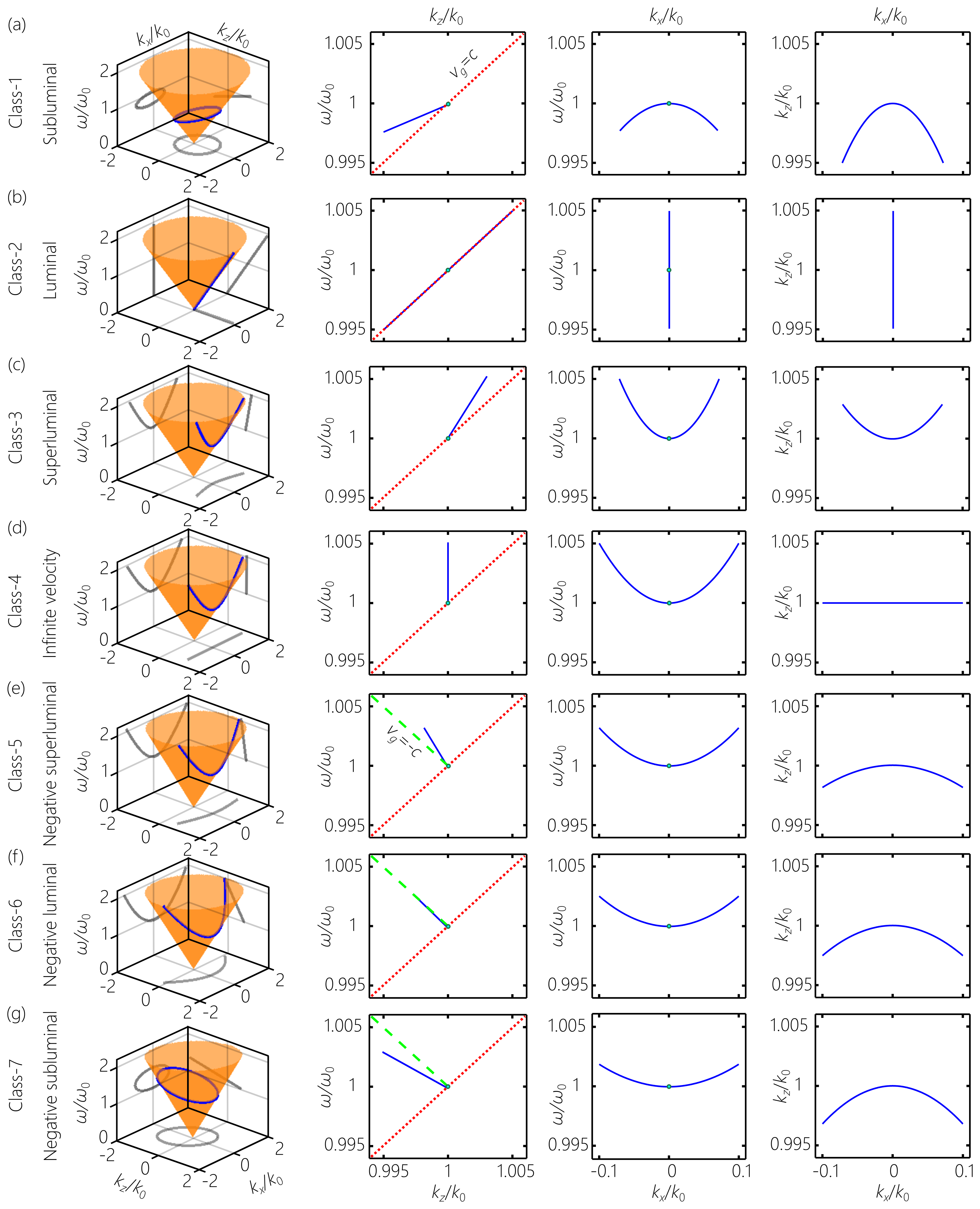}
  \end{center}
  \caption{Spatio-temporal spectra of \textit{baseband} ST wave packets. (a) Class-1, positive subluminal; $\theta\!=\!\tfrac{\pi}{6}$. (b) Class-2, positive luminal plane-wave pulse, $\theta\!=\!\tfrac{\pi}{4}$. (c) Class-3, positive superluminal; $\theta\!=\!\tfrac{\pi}{3}$. (d) Class-4, infinite-$v_{\mathrm{g}}$; $\theta\!=\!\tfrac{\pi}{2}$. (e) Class-5, negative superluminal; $\theta\!=\!\tfrac{2\pi}{3}$. (f) Class-6, negative luminal; $\theta\!=\!\tfrac{3\pi}{4}$. (g) Class-7, negative subluminal; $\theta\!=\!\tfrac{5\pi}{6}$. The projections are idealized 1D geometric curves. In practice, a finite spectral uncertainty is introduced into the width of these curves. In all cases we plot to $|k_{x}|\!=\!0.1k_{\mathrm{o}}$. The dotted red line corresponds to $v_{\mathrm{g}}\!=\!c$, and the dashed green line to $v_{\mathrm{g}}\!=\!-c$.}
  \label{Fig:Projections3-4-5-6-7}
\end{figure*}

We make use of three criteria to classify ST wave packets: (1) \textit{Group velocity magnitude} $v_{\mathrm{g}}$: The group velocity $v_{\mathrm{g}}\!=\!\tfrac{\partial\omega}{\partial k_{z}}\!=\!c\tan{\theta}$ may in principle be subluminal $v_{\mathrm{g}}\!<\!c$, luminal $v_{\mathrm{g}}\!=\!c$, or superluminal $v_{\mathrm{g}}\!>\!c$. (2) \textit{Direction of $v_{\mathrm{g}}$}: The ST wave packet may propagate forward \textit{away} from the source $v_{\mathrm{g}}\!>\!0$ or backward \textit{towards} it $v_{\mathrm{g}}\!<\!0$. (3) \textit{Baseband or sideband spatial spectra}: If low spatial frequencies are allowed, we refer to the ST wave packet as \textit{baseband}, otherwise we use the term \textit{sideband} ST wave packet, which correspond to the planes $\mathcal{P}_{\mathrm{b}}$ and $\mathcal{P}_{\mathrm{s}}$, respectively. A third option is X-waves lying in $\mathcal{P}_{\mathrm{o}}$, which are baseband in principle but sideband in practice. In addition, we impose two general restrictions: (1) to avoid evanescent components, we consider only plane-wave contributions lying \textit{on} the surface of the light-cone; and (2) we consider only \textit{positive} (forward-propagating) values of $k_{z}$ to ensure compatibility with causal excitation.

We list in Table~\ref{Table:Classification} the 10 physically realizable and unique classes of ST wave packets. Class-1 comprises baseband positive subluminal ST wave packets $v_{\mathrm{g}}\!<\!c$; Class-2 is that of baseband positive luminal ST wave packets $v_{\mathrm{g}}\!=\!c$, which is a degenerate case with $k_{x}\!=\!0$ (pulsed plane waves); Class-3 is that of baseband positive superluminal ST wave packets $v_{\mathrm{g}}\!>\!c$; Class-4 is that of baseband ST wave packets formally with infinite $v_{\mathrm{g}}$; Class-5 is that of baseband ST wave packets with negative superluminal group velocity $|v_{\mathrm{g}}|\!>\!c$; Class-6 is that of baseband negative luminal ST wave packets, $v_{\mathrm{g}}\!=\!-c$; and Class-7 is that of baseband negative subluminal ST wave packets $|v_{\mathrm{g}}|\!<\!c$. The remaining classes are sideband ST wave packets: Class-8 a positive luminal class $v_{\mathrm{g}}\!=\!c$, and Class-9 a positive superluminal class $v_{\mathrm{g}}\!>\!c$. Finally, Class-10 is that of positive superluminal X-waves \cite{Lu92IEEEa}, which we associate with \textit{sideband} ST wave packets because they share the same angular range $\tfrac{\pi}{4}\!<\!\theta\!<\!\tfrac{\pi}{2}$.

Our classification admits in principle $3\times2\times3\!=\!18$ classes of ST wave packets, of which 9 are eliminated on physical grounds: Luminal X-waves coincide with Class-2; subluminal X-waves cannot be synthesized with real $k_{z}$; negative subluminal, luminal, and superluminal X-waves all correspond to $k_{z}\!<\!0$; postive subluminal sideband ST wave packets with Class-1 (see below); and negative subluminal, luminal, and superluminal sideband ST wave packets require $k_{z}\!<\!0$. All these can be seen clearly from the geometric models in Fig.~\ref{Fig:Projections3-4-5-6-7} and Fig.~\ref{Fig:Projections9-9-10}. Note that Class-4 of nominally infinite-$v_{\mathrm{g}}$ is listed separately for its interesting `time-diffraction' properties \cite{Kondakci16OE,Parker16OE,Porras17OL,Kondakci18PRL} and is only a limit separating Class-3 and Class-5.  

Although $v_{\mathrm{g}}$ can take on arbitrary values \cite{Turunen10PO,Salo01JOA,Garay16AO} (determined by $\theta$ \cite{Kondakci18unpub}), there is no violation of special relativity; $v_{\mathrm{g}}$ is not the information velocity \cite{Smith70AJP}, only the velocity of the peak of the wave packet \cite{SaariPRA18}. ST wave packets can be viewed as an instance of the so-called `scissors effect', whereupon the point of intersection of two very long blades of a pair of scissors can move at a considerably higher speed than that of its ears but without conveying information; e.g., the apparent faster-than-light motion of a spotlight on a faraway screen \cite{Salmon16Book}. Likewise, a negative group $v_{\mathrm{g}}$ corresponds to the closing of shears whose blades are connected at their far end: closing the shears leads the intersection point to move backwards.

\subsection{Baseband ST wave packets}

\begin{table*}[htb]\label{Table:Baseband}
  \centering
  \caption{Characteristics of baseband ST wave packets.}
  \begin{tabular}{p{0.4in}p{0.7in}p{1.6in}p{1.6in}p{0.5in}p{0.7in}p{0.7in}} \hline\hline
  &&&&&\\
     Class & $\theta$ & $(k_{x},\tfrac{\omega}{c})$ & $(k_{x},k_{z})$ & $\omega_{\mathrm{o}}$ & $|k_{x}^{\mathrm{max}}|/k_{\mathrm{o}}$ & $(\Delta\omega)^{\mathrm{max}}/\omega_{\mathrm{o}}$\\&&&&&\\ \hline\hline

&&&&&\\
(1) &
$0\!<\!\theta\!<\!\frac{\pi}{4}$ & $\frac{1}{k_{1}^{2}}\left(\frac{\omega}{c}-k_{2}\right)^{2}+\frac{k_{x}^{2}}{k_{3}^{2}}=1$ &
$\frac{1}{k_{2}^{2}}\left(k_{z}-k_{1}\right)^{2}+\frac{k_{x}^{2}}{k_{3}^{2}}=1$ &
max. &
$k_{3}/k_{\mathrm{o}}$ & 
$\tan{\theta}$
\\
&&&&&\\\hline
&&&&&\\

(2) &
$\theta\!=\!\frac{\pi}{4}$ &
$k_{x}\!=\!0$ &
$k_{x}\!=\!0$ & 
center &
--- &
---\\
&&&&&\\\hline

&&&&&\\
(3) & $\frac{\pi}{4}\!<\!\theta\!<\!\frac{\pi}{2}$ &
$\frac{1}{k_{1}^{2}}\left(\frac{\omega}{c}-k_{2}\right)^{2}-\frac{k_{x}^{2}}{k_{3}^{2}}=1$ &
$\frac{1}{k_{2}^{2}}\left(k_{z}-k_{1}\right)^{2}-\frac{k_{x}^{2}}{k_{3}^{2}}=1$ &
min. &
--- & 
---\\
&&&&&\\ \hline

&&&&&\\
(4) &
$\theta\!=\!\frac{\pi}{2}$ &
$(\frac{\omega}{c})^{2}-k_{x}^{2}=k_{\mathrm{o}}^{2}$ &
$k_{z}\!=\!k_{\mathrm{o}}$ & 
min. &
--- &
---\\
&&&&&\\\hline

&&&&&\\
(5) &
$\frac{\pi}{2}\!<\!\theta\!<\!\frac{3\pi}{4}$ &
$\frac{1}{k_{1}^{2}}\left(\frac{\omega}{c}+k_{2}\right)^{2}-\frac{k_{x}^{2}}{k_{3}^{2}}=1$ &
$\frac{1}{k_{2}^{2}}\left(k_{z}+k_{1}\right)^{2}-\frac{k_{x}^{2}}{k_{3}^{2}}=1$ & 
min. &
$1-\tan{\theta}$ &
$|\tan{\theta}|$\\
&&&&&\\\hline

&&&&&\\
(6) &
$\theta\!=\!\frac{3\pi}{4}$ &
$\frac{\omega}{c}=\frac{1}{4k_{\mathrm{o}}}k_{x}^{2}+k_{\mathrm{o}}$ &
$k_{z}=-\frac{1}{4k_{\mathrm{o}}}k_{x}^{2}+k_{\mathrm{o}}$ & 
min. &
$2$ &
$1$\\
&&&&&\\\hline

&&&&&\\
(7) &
$\frac{3\pi}{4}\!<\!\theta\!<\!\pi$ &
$\frac{1}{k_{1}^{2}}\left(\frac{\omega}{c}-k_{2}\right)^{2}+\frac{k_{x}^{2}}{k_{3}^{2}}=1$ &
$\frac{1}{k_{2}^{2}}\left(k_{z}+k_{1}\right)^{2}+\frac{k_{x}^{2}}{k_{3}^{2}}=1$ & 
min. &
$1-\tan{\theta}$ &
$|\tan{\theta}|$\\
&&&&&\\
\hline\hline
\end{tabular}
\end{table*}

We plot in Fig.~\ref{Fig:Projections3-4-5-6-7} the projections of the 7 classes of \textit{baseband} ST wave packets with increasing $\theta$ and provide the equation in Table~\ref{Table:Baseband}.

\begin{figure}[t!]
  \begin{center}
 \includegraphics[width=8.6cm]{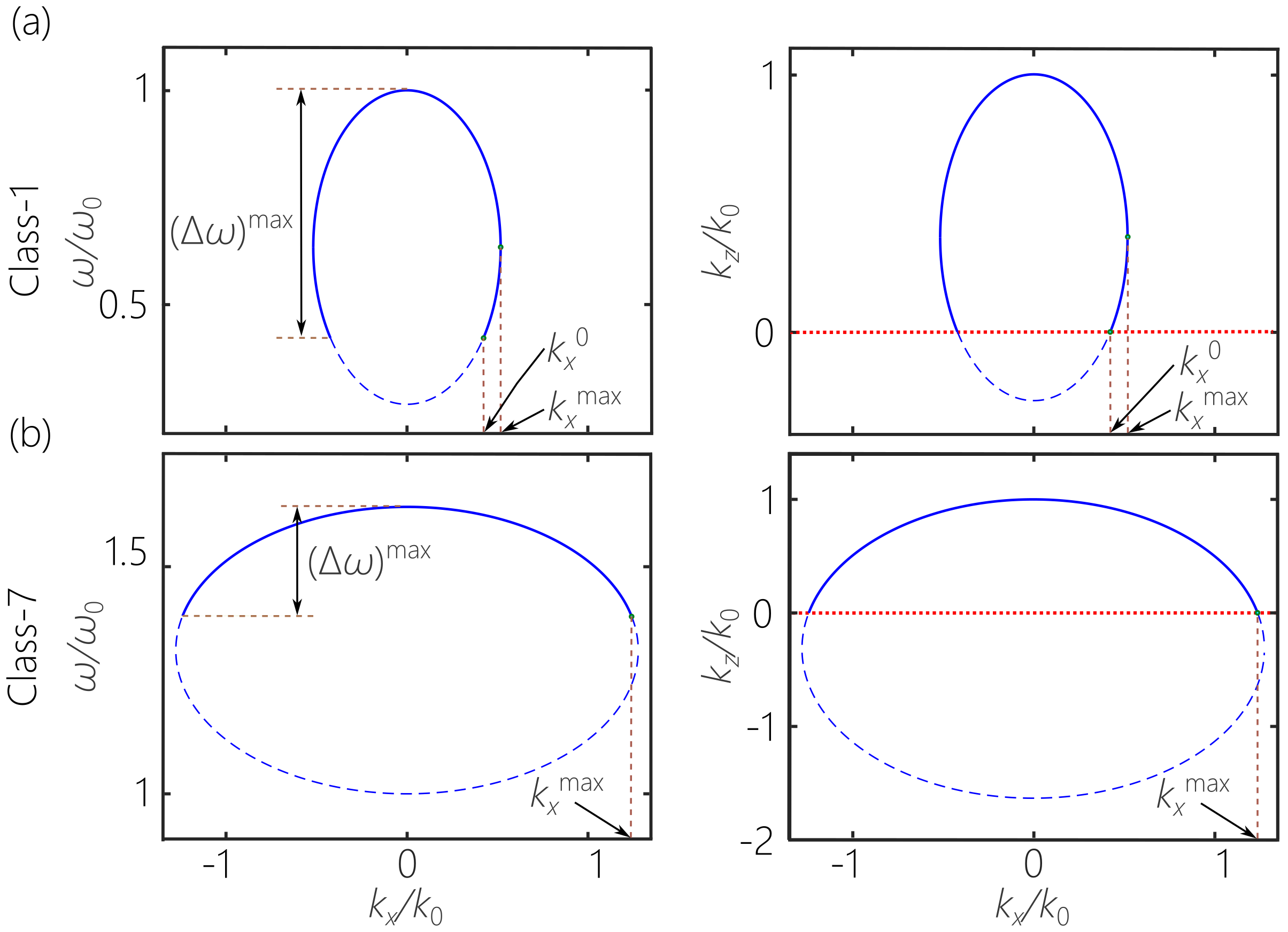}
  \end{center}
  \caption{Projections on the $(k_{x},\tfrac{\omega}{c})$ and $(k_{x},k_{z})$ planes for (a) Class-1  with $\theta\!=\!30^{\circ}$ and (b) Class-7 with $\theta\!=\!166.5^{\circ}$. Dashed portions are excluded where $k_{z}\!<\!0$. In (a) we highlight the range $k_{x}^{0}\!<\!k_{x}\!<\!k_{x}^{\mathrm{max}}$ where the values of $k_{x}$ are in a two-to-one relationship with $\omega$.}
  \label{Fig:ellipses}
\end{figure}

\textit{Class-1:} In the positive subluminal range $0\!<\!\theta\!<\!\tfrac{\pi}{4}$ and $\omega_{\mathrm{o}}$ is the \textit{maximum} temporal frequency [Fig.~\ref{Fig:Projections3-4-5-6-7}(a)]. The projection onto the $(k_{x},\tfrac{\omega}{c})$-plane is an ellipse, but becomes a circle at $\tan{\theta}\!=\!1/\!\sqrt{2}$, whereupon $k_{1}\!=\!k_{3}\!=\!k_{\mathrm{o}}/(1\!+\!\sqrt{2})$; Fig.~\ref{Fig:k1k2k3}(a). A critical consequence of the constraint $k_{z}\!\geq\!0$ is the existence of a maximum value of $k_{x}$, $|k_{x}^{\mathrm{max}}|\!=\!k_{3}$, corresponding to $\tfrac{\omega}{c}\!=\!k_{2}$, and thus a minimum achievable spatial and temporal widths. Further decrease in $k_{z}$ is associated with a \textit{decrease} in $k_{x}$ below $k_{x}^{\mathrm{max}}$ [Fig.~\ref{Fig:ellipses}(a)], and reaching the limit $k_{z}\!=\!0$ corresponds to $k_{x}^{0}\!=\!k_{\mathrm{o}}(1-\tan{\theta})$.  Unexpectedly, there is a two-to-one relationship between $\omega$ and $|k_{x}|$ in the range extending between $k_{x}^{\mathrm{max}}$ and $k_{x}^{0}$, a feature that is unique to Class-1. In all other classes the relationship between $|k_{x}|$ and $\omega$ is strictly one-to-one. To the best of our knowledge, Class-1 has not been realized to date except in Refs.~\cite{Kondakci17NP,Kondakci18unpub}.

\textit{Class-2:} The case of $\theta\!=\!\tfrac{\pi}{4}$ ($v_{\mathrm{g}}\!=\!c$) corresponds the tangent to the light-cone at $k_{x}\!=\!0$ [Fig.~\ref{Fig:Projections3-4-5-6-7}(b)]. Class-2 ST wave packets are pulsed \textit{plane waves} that do not disperse in free space and naturally do not diffract \cite{SalehBook07}. 

\textit{Class-3:} For positive superluminal propagation $v_{\mathrm{g}}\!>\!c$, $\mathcal{P}_{\mathrm{b}}(\theta)$ is tilted in the range $\tfrac{\pi}{4}\!<\theta\!<\!\tfrac{\pi}{2}$ and intersects with the light-cone in a hyperbola, where $\omega_{\mathrm{o}}$ is now the \textit{minimum} temporal frequency; see Fig.~\ref{Fig:Projections3-4-5-6-7}(c). An approach to synthesizing such wave packets was proposed in Ref.~\cite{Valtna07OC}. To the best of our knowledge, Class-3 has \textit{not} been realized to date except in Refs.~\cite{Kondakci17NP,Kondakci18unpub}.

In general, there are no restrictions on the maximum bandwidth for positive superluminal ST wave packets (Class-3 through Class-6), and hence no lower limit on the transverse spatial width or pulse width, except for experimental limits on the broad temporal bandwidth.

\textit{Class-4:} Although Class-4 (infinite-$v_{\mathrm{g}}$ at $\theta\!=\!\tfrac{\pi}{2}$) is only the boundary between the regimes of positive- and negative-$v_{\mathrm{g}}$, we highlight it for its unique properties. The plane-wave expansion in Eq.~\ref{Eq:STExpansion} simplifies for Class-3, $\psi(x,z,t)\!=\!\int\!dk_{x}\widetilde{\psi}(k_{x})e^{i\{k_{x}x-(\omega-\omega_{\mathrm{o}})t\}}$, such that $t$ here replaces $z$ in a traditional monochromatic beam. Consequently, the diffractive behavior usually observed with axial propagation is now observed in \textit{time}. This phenomenon appears to have been identified initially in Ref.~\cite{Longhi04OE} and labeled `temporal diffraction', was studied theoretically in \cite{Kondakci16OE}, and more recently investigated theoretically in \cite{Porras17OL,PorrasPRA18} where it has been labeled `time diffraction', and experimentally in \cite{Kondakci18PRL}. In practice, $v_{\mathrm{g}}$ is finite, albeit large, due to the unavoidable finite spectral uncertainty $\delta\omega$ in identifying $k_{x}$ with $\omega$ \cite{Kondakci17NP}.

\textit{Class-5:} When $\theta\!>\!\tfrac{\pi}{2}$, the group velocity is negative. In the range $\tfrac{\pi}{2}\!<\!\theta\!<\!\tfrac{3\pi}{4}$, Class-5 is superluminal $|v_{\mathrm{g}}|\!>\!c$, thus traveling \textit{backwards} in the negative-$z$ direction. The intersection of the light-cone with $\mathcal{P}_{\mathrm{b}}(\theta)$ is a hyperbola [Fig.~\ref{Fig:Projections3-4-5-6-7}(e)]. Class-5 is characterized by upper limits on the spatial bandwidth as a result of the restriction $k_{z}\!\geq\!0$, which in turn imposes a maximum temporal bandwidth of more than one octave [Table~\ref{Table:Baseband}]. That is, for a given $\theta$, Class-5 has \textit{minimum} achievable spatial and temporal widths. A method for synthesizing Class-5 has been proposed in \cite{Zapata06OL} but has only been realized in \cite{Kondakci18unpub}.

\textit{Class-6:} At the negative luminal limit $\theta\!=\!\tfrac{3\pi}{4}$, $v_{\mathrm{g}}\!=\!-c$, the intersection of the light-cone with $\mathcal{P}_{\mathrm{b}}(\frac{3\pi}{4})$ is a parabola with $\omega_{\mathrm{o}}$ corresponding to the \textit{minimum} temporal frequency [Fig.~\ref{Fig:Projections3-4-5-6-7}(f)]. There is a \textit{maximal} allowable spatial frequency, and the associated upper limit on the temporal bandwidth corresponds to a full octave [Table~\ref{Table:Baseband}], resulting in lower limits on the spatial and temporal widths. To the best of our knowledge, Class-6 has \textit{not} been realized experimentally to date.

\textit{Class-7:} When $\tfrac{3\pi}{4}\!<\!\theta\!<\!\pi$ we reach a regime where the magnitude of the negative $v_{\mathrm{g}}$ is subluminal $|v_{\mathrm{g}}|\!<\!c$. The intersection of the light-cone with $\mathcal{P}_{\mathrm{b}}(\theta)$ is an ellipse and $\omega_{\mathrm{o}}$ is the \textit{maximum} temporal frequency [Fig.~\ref{Fig:Projections3-4-5-6-7}(g)]. The projected ellipse in the $(k_{x},\tfrac{\omega}{c})$-plane becomes a circle at $\tan{\theta}\!=\!-1/\!\sqrt{2}$, whereupon $k_{1}\!=\!k_{3}$ [Fig.~\ref{Fig:k1k2k3}(a)]. As in the previous two classes, there is an \textit{upper} limit of the achievable spatial bandwidth and the corresponding temporal bandwidth is less than one octave [Table~\ref{Table:Baseband}]. In contrast to Class-1, the one-to-one correspondence between $|k_{x}|$ and $\omega$ is maintained in Class-7 as shown in Fig.~\ref{Fig:ellipses}(b). Class-7 has \textit{not} been realized experimentally to date.

\subsection{Sideband ST wave packets and X-waves}
\begin{figure*}[t!]
  \begin{center}
  \includegraphics[width=14.5cm]{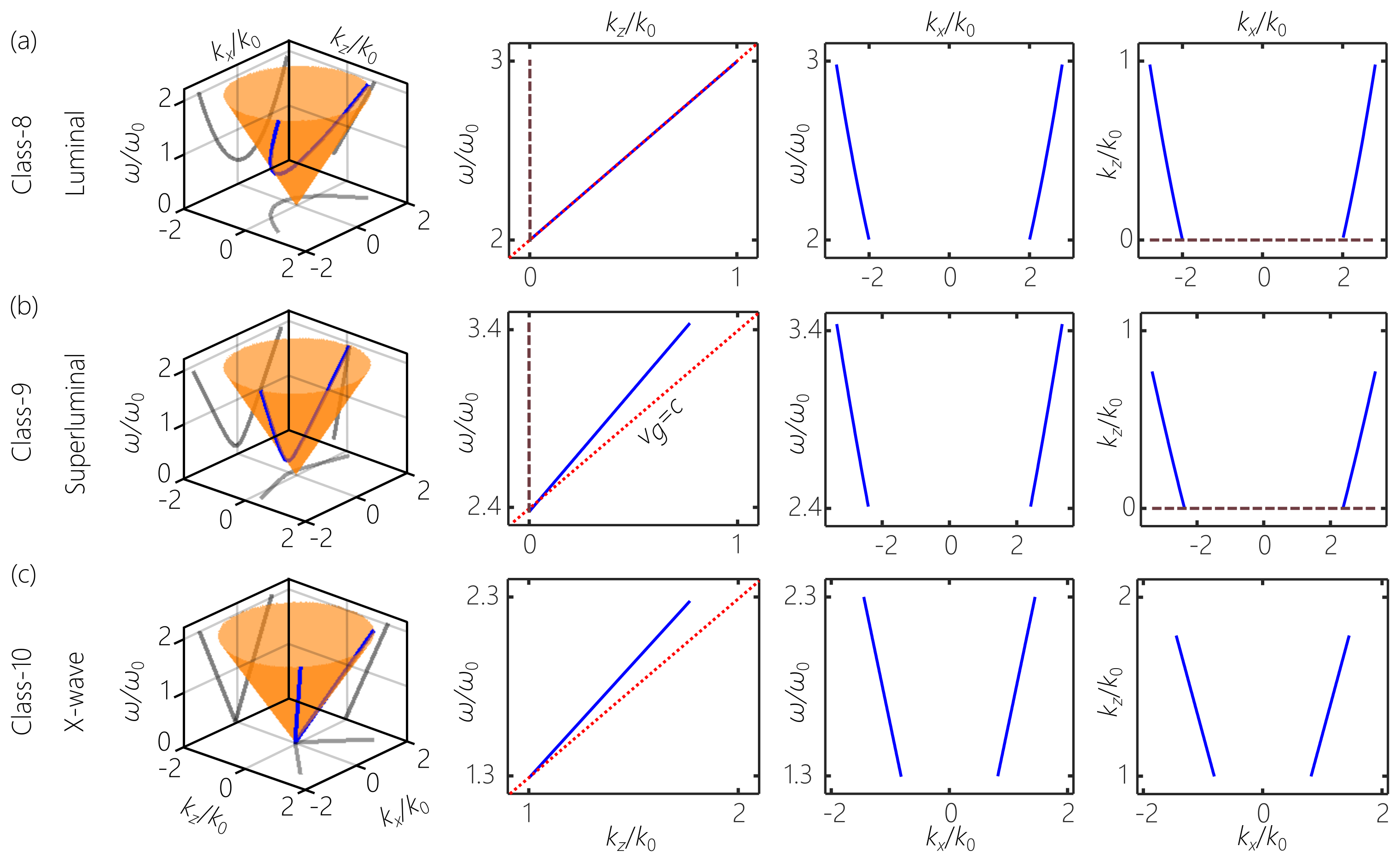}
  \end{center}
  \caption{Spatio-temporal spectra of \textit{sideband} ST wave packets. (a) Class-8, positive luminal; $\theta\!=\!\tfrac{\pi}{4}$. (b) Class-9, positive superluminal; $\theta\!=\!0.3\pi$. (c) Class-10, positive superluminal (X-waves); $\theta\!=\!0.289\pi$. The dotted red line corresponds to $v_{\mathrm{g}}\!=\!c$ (but is not necessarily the light-line, and is provided as a guide for the eye), and the dashed black line to $k_{z}\!=\!0$.}
  \label{Fig:Projections9-9-10}
\end{figure*}

We now move on to \textit{sideband} ST wave packets [Fig.~\ref{Fig:Projections9-9-10}(a,b)]; see Table~\ref{Table:Sideband} for details. The low-frequency range in the vicinity of $k_{x}\!=\!0$ is excluded where $k_{z}\!<\!0$. The \textit{minimum} allowed value of $k_{x}$ associated with $k_{z}\!=\!0$. Beyond the minimum values, there are no \textit{upper} limits on the spatial or temporal frequencies. Therefore, arbitrary spatial and temporal widths are achievable.

In the subluminal range $0\!<\!\theta\!<\!\tfrac{\pi}{4}$, the intersection has the form of an ellipse that is \textit{not} unique, and in fact coincides with Class-1 after making the substitution
$k_{\mathrm{o}}\rightarrow k_{\mathrm{o}}\tfrac{1+\tan{\theta}}{1-\tan{\theta}}$. We thus do \textit{not} identify this case as a unique class. 

\textit{Class-8:} In the positive luminal case $\theta\!=\!\tfrac{\pi}{4}$ ($v_{\mathrm{g}}\!=\!c$), $\mathcal{P}_{\mathrm{s}}(\tfrac{\pi}{4})$ intersects the light-cone in a parabola. The well-known example of Brittingham's FWM \cite{Brittingham83JAP} belongs to Class-8. An optical arrangement for synthesizing such wave packets was proposed in \cite{Reivelt00JOSAA} and initial experimental results were reported in \cite{Reivelt02PRE}.

\textit{Class-9:} In the positive superluminal range $\tfrac{\pi}{4}\!<\!\theta\!<\!\tfrac{\pi}{2}$ ($v_{\mathrm{g}}\!>\!c$), $\mathcal{P}_{\mathrm{s}}(\theta)$ intersects with the light-cone in a hyperbola. To the best of our knowledge, Class-9 has \textit{not} been synthesized experimentally to date. In contrast to the hypothetical subluminal sideband ST wave packets, Class-8 and Class-9 cannot be transformed into baseband counterparts. 

\begin{table*}[htb]\label{Table:Sideband}
  \centering
  \caption{Characteristics of sideband ST wave packets.}
  \begin{tabular}{p{0.4in}p{0.7in}p{1.6in}p{1.6in}p{0.5in}p{0.7in}p{0.7in}} \hline\hline
  &&&&&\\
     Class & $\theta$ & $(k_{x},\tfrac{\omega}{c})$ & $(k_{x},k_{z})$ & $\omega_{\mathrm{o}}$ & $|k_{x}^{\mathrm{min}}|/k_{\mathrm{o}}$ & $(\Delta\omega)^{\mathrm{min}}/\omega_{\mathrm{o}}$\\&&&&&\\ \hline\hline

&&&&&\\
(8) &
$\theta\!=\!\frac{\pi}{4}$ & $\frac{\omega}{c}=\frac{1}{4k_{\mathrm{o}}}k_{x}^{2}+k_{\mathrm{o}}$ &
$k_{z}=\frac{1}{4k_{\mathrm{o}}}k_{x}^{2}-k_{\mathrm{o}}$ &
min. &
$2$ & 
$2$\\
&&&&&\\\hline
&&&&&\\

(9) &
$\frac{\pi}{4}\!<\!\theta\!<\!\frac{\pi}{2}$ &
$\frac{1}{k_{1}'^{2}}\left(\frac{\omega}{c}+k_{2}'\right)^{2}-\frac{k_{x}^{2}}{k_{3}'^{2}}=1$ &
$\frac{1}{k_{2}'^{2}}\left(k_{z}+k_{1}'\right)^{2}-\frac{k_{x}^{2}}{k_{3}'^{2}}=1$ &
min. &
$1+\tan{\theta}$ &
$1+\tan{\theta}$\\
&&&&&\\\hline

&&&&&\\
(10) & $\frac{\pi}{4}\!<\!\theta\!<\!\frac{\pi}{2}$ &
$\frac{\omega}{c}=\frac{k_{x}\tan{\theta}}{\sqrt{\tan^{2}{\theta}-1}}$ &
$\frac{\omega}{c}=\frac{k_{x}}{\sqrt{\tan^{2}{\theta}-1}}$ &
--- &
0 & 
0\\
&&&&&\\ 
\hline\hline

\end{tabular}
\end{table*}

\textit{Class-10:} X-waves lie at the intersection of $\mathcal{P}_{\mathrm{o}}(\theta)$ with the light-cone in the superluminal range $\tfrac{\pi}{4}\!<\theta\!<\!\tfrac{\pi}{2}$ [Fig.~\ref{Fig:Projections9-9-10}(c)]. Whereas previous experiments on synthesizing optical X-waves realized 2D spatial profiles \cite{Saari97PRL}, there have been no demonstrations to date of 1D X-waves in Class-10.

\begin{figure}[t!]
  \begin{center}
  \includegraphics[width=8.5cm]{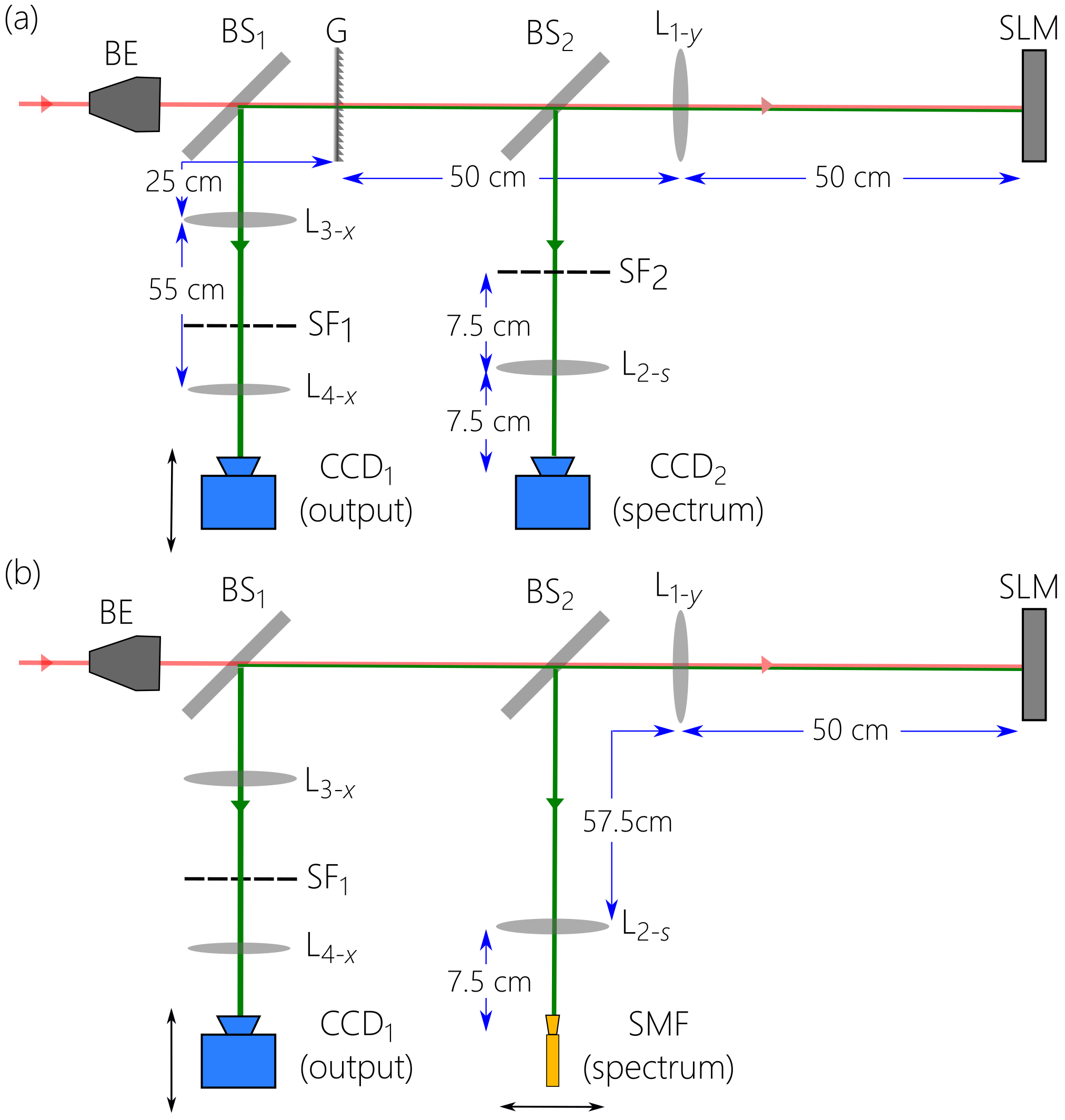}
  \end{center}
  \caption{Schematic depiction of the experimental setup to synthesize and characterize (a) ST wave packets; (b) monochromatic and traditional pulsed beams. BE: Beam expander; BS$_1$, BS$_2$: beam splitters; G: diffraction grating; CCD$_1$, CCD$_2$: CCD cameras; SLM: spatial light modulator; SMF: single mode fiber; SF$_1$, SF$_2$: spatial filters; L$_{1-y}$, L$_{3-x}$, L$_{4-x}$: cylindrical lenses; L$_{2-\mathrm{s}}$: spherical lens. The focal lengths of all the lenses and their separations are given in the figure.}
  \label{Fig:Setup}
\end{figure}

\section{Generation of space-time wave packets}\label{Sec:Experiment}

\subsection{Experimental setup}

Our experimental strategy is based on ultrafast pulse-shaping via spectral phase modulation \cite{Weiner00RSI,Weiner09Book}, but utilizes a 2D phase modulation scheme to associate a spatial frequency $k_{x}$ with each temporal frequency or wavelength $\lambda$ \cite{Koehl98OC,Feurer02OL,Tanab05AO,Zhu05OE,Sussman08PRA,He14OE}. This scheme allows for programming an arbitrary correlation function between $\lambda$ and $k_{x}$ and thus the synthesis of \textit{any} class of ST wave packets. Figure~\ref{Fig:Setup}(a) illustrates the setup utilized in synthesizing ST wave packets. A horizontally polarized large-area femtosecond pulsed beam from a Ti:Sa laser (Tsunami, Spectra Physics; bandwidth of $\sim\!8.5$~nm centered on a wavelength of $\sim\!800$~nm) is incident at an angle of $68.5^{\circ}$ on a reflective diffraction grating G (1200~lines/mm and area $25\!\times\!25$~mm$^{2}$; Newport 10HG1200-800-1). The second diffraction order is collimated by a cylindrical lens L$_{1-y}$ (focal length $f\!=\!50$~cm) oriented along the $y$-direction in a $2f$ configuration such that a temporal bandwidth of $\sim2.1$~nm is spread over the transverse width $\sim\!16$~mm of a reflecting phase-only SLM (Hamamatsu X10468-02), which modulates the impinging wave front with a 2D spatial phase $\Phi(x,y)$ to enforce a programmable one-to-one correspondence between $|k_{x}|$ and $\lambda$. The phase-modulated wavefront is retro-reflected back through the lens L$_{1-y}$ to G, whereupon the ST wave packet is produced in the form of a light sheet.

For sake of comparison, we also synthesize a monochromatic beam and a traditional pulsed beam with \textit{uncoupled} spatial and temporal spectra using the modified setup shown in Fig.~\ref{Fig:Setup}(b). The grating G is removed, such that the beam is incident on the SLM without first spreading the temporal spectrum. The spatial frequencies are then assigned to the \textit{whole} temporal spectrum. The monochromatic beam is produced from a CW laser diode (Thorlabs, CPS808S, tunable in the wavelength range $804-806$~nm, bandwidth $<\!0.2$~nm). For the pulsed beam, we make use of the same Ti:Sa laser as input. To maximize the utilization of the SLM, we design $\Phi(x,y)$ to reduced $k_{x}$ by a factor of $10\times$ for baseband ST wave packets (a factor of $100\times$ for sideband ST wave packets because of the higher spatial frequencies involved) and the retrieve the intended spatial spectrum via two cylindrical lenses L$_{3-x}$ ($f\!=\!50$~cm) and L$_{4-x}$ ($f\!=\!5$~cm) oriented along the $x$-direction; see Fig.~\ref{Fig:Setup}. In the Fourier plane between the lenses L$_{3-x}$ and L$_{4-x}$, a spatial filter (SF$_1$) in the form of a thin wire rejects low spatial frequencies to remove the unwanted DC components from the finite diffraction efficiency of the SLM. 

\begin{figure}[t!]
  \begin{center}
  \includegraphics[width=8.6cm]{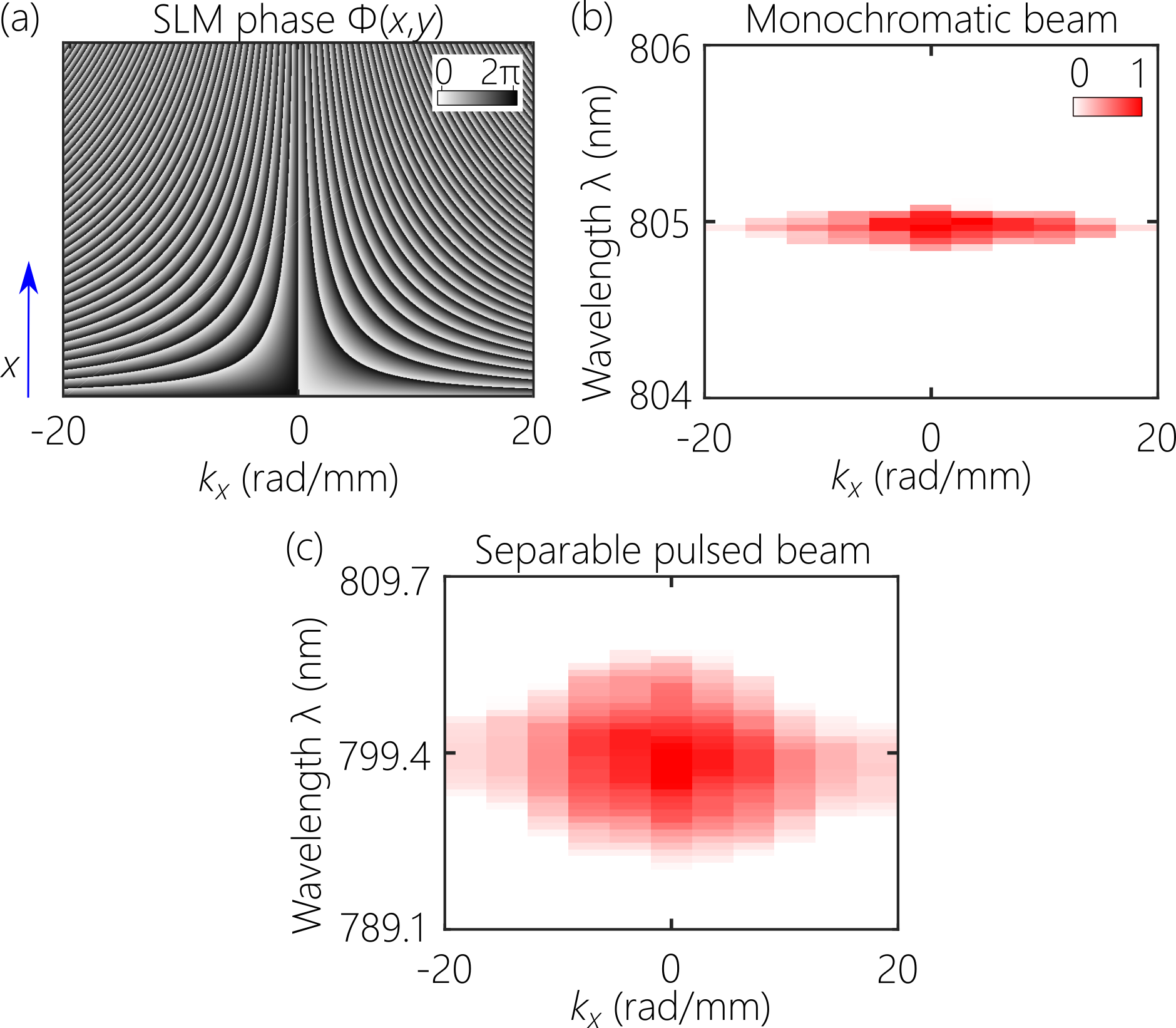}
  \end{center}
  \caption{(a) Implemented 2D SLM phase distribution $\Phi(x,y)$ to produce monochromatic and separable pulsed beams; $\Delta k_{x}\!=\!20$~rad/mm. (b) Measured spatio-temporal spectrum $|\tilde{E}(k_{x},\lambda)|^{2}$ for a monochromatic beam, corresponding to Fig.~\ref{Fig:Projections1-2}(a); and (c) for a separable pulsed beam having $\Delta\lambda\!=\!8.5$~nm, corresponding to Fig.~\ref{Fig:Projections1-2}(b).}
  \label{Fig:Spectral1-2}
\end{figure}

The axial evolution of the time-averaged intensity profile $I(x,y,z)\!=\!\int\!dt|E(x,y,z,t)|^{2}$ is recorded by scanning a CCD camera along $z$ (CCD$_1$; the Imaging Source, DMK 33UX178). The uniform intensity along $y$ ($\sim\!25$~mm) is averaged over a segment of width $\Delta y\!\sim\!1.32$~mm, $I(x,z)\!=\!\int_{\Delta y}\!dyI(x,y,z)$. To characterize the spatio-temporal spectrum of the ST wave packets $|\widetilde{\psi}(k_{x},\lambda)|^{2}$, a portion of the beam retro-reflected from the SLM is directed by a beam splitter (BS$_2$) through a spherical lens L$_{2-\mathrm{s}}$ ($f\!=\!7.5$~cm) to a CCD camera (CCD$_2$; The Imaging Source, DMK 72AUC02). The spatial frequencies along the $x$-direction undergo a Fourier transform via $L_{1-\mathrm{s}}$ and are unaffected by $L_{1-y}$. The unwanted DC spectral components is removed by a spatial filter SF$_{2}$ placed in the Fourier plane between the lenses L$_{1-y}$ and L$_{2-\mathrm{s}}$ [Fig.~\ref{Fig:Setup}(a)]. This spatial filter is removed when synthesizing Class-2 wave packets that correspond to a plane wave. The temporal spectral resolution of the measurement is limited by the pixel size of CCD$_2$, and an accurate estimate of $\delta\lambda$ is obtained with an optical spectrum analyzer (Advantest Q8384). When a monochromatic or separable pulsed beam is synthesized, we use the modified system shown in Fig.~\ref{Fig:Setup}(b), where CCD$_{2}$ is replaced with a single-mode fiber (delivering to an optical spectrum analyzer) that is scanned in the focal plane of L$_{2-\mathrm{s}}$.

\subsection{Measurements of monochromatic and separable pulsed beams}

The SLM phase pattern $\Phi(x,y)$ utilized is shown in Fig.~\ref{Fig:Spectral1-2}(a). The right and left halves correspond to positive- and negative-valued $k_{x}$. The spatio-temporal spectrum ideally is $\tilde{E}(k_{x},\lambda)\!=\!\widetilde{E}_{x}(k_{x})\delta(\lambda-\lambda_{\mathrm{o}})$ for the monochromatic beam ($\lambda_{\mathrm{o}}$ is a fixed wavelength), and $\tilde{E}(k_{x},\lambda)\!=\!\widetilde{E}_{x}(k_{x})\widetilde{E}_{t}(\lambda)$ for the traditional pulsed beam, where $\widetilde{E}_{x}$ and $\widetilde{E}_{t}$ are the spatial and temporal spectra, respectively. Each is separable in the spatial and temporal degrees of freedom. The measured spatio-temporal spectrum obtained via the configuration in Fig.~\ref{Fig:Setup}(b) is plotted in Fig.~\ref{Fig:Spectral1-2}(b) for the monochrmatic beam and in Fig.~\ref{Fig:Spectral1-2}(c) for the pulsed beam, confirming their separability. Neither reveals any spatio-temporal spectral correlations, and we thus expect that both will undergo diffractive spreading, as we confirm below.

\begin{figure}
  \begin{center}
  \includegraphics[width=8.5cm]{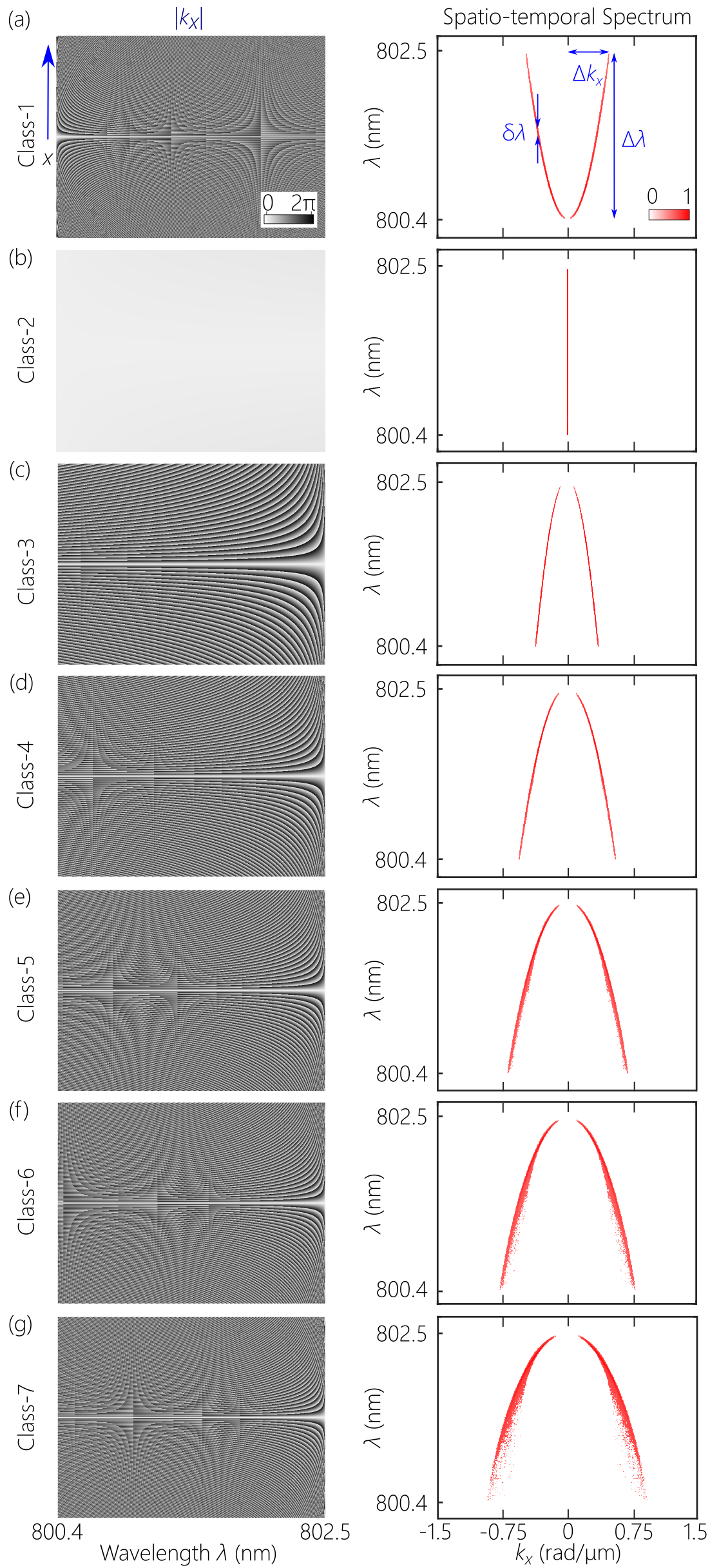}
  \end{center}
  \caption{Implemented 2D SLM phase distributions $\Phi(x,y)$ (left column) and the \textit{measured} spatio-temporal spectra $|\tilde{E}(k_{x},\lambda)|^{2}$ (right column) for baseband ST wave packets, Class-1 through Class-7, as listed in Table~\ref{Table:Classification} and depicted in Fig.~\ref{Fig:Projections3-4-5-6-7}. The temporal bandwidth for all classes is $\Delta\lambda\!=\!2.1$~nm. (a) Class-1 with $\theta\!=\!30^{\circ}$ and $\Delta k_{x}\!=\!0.48$~rad/$\mu$m; (b) Class-2 with $\theta\!=\!45^{\circ}$ and $\Delta k_{x}\!=\!0$~rad/$\mu$m; (c) Class-3 with $\theta\!=\!60^{\circ}$ and $\Delta k_{x}\!=\!0.36$~rad/$\mu$m; (d) Class-4 with $\theta\!=\!90^{\circ}$ and $\Delta k_{x}\!=\!0.56$~rad/$\mu$m; (e) Class-5 with $\theta\!=\!120^{\circ}$ and $\Delta k_{x}\!=\!0.7$~rad/$\mu$m; (f) Class-6 with $\theta\!=\!135^{\circ}$ and $\Delta k_{x}\!=\!0.79$~rad/$\mu$m; and (g) Class-7 with $\theta\!=\!150^{\circ}$ and $\Delta k_{x}\!=\!0.93$~rad/$\mu$m.}
  \label{Fig:Spectral3-4-5-6-7}
\end{figure}

\subsection{Measurements of baseband ST wave packets}

We now move on to the synthesis of \textit{baseband} ST wave packets, Class-1 through Class-7. The 2D phase distribution utilized at a fixed temporal bandwidth of $\Delta\lambda\!\sim\!2.1$~nm and the corresponding measured spatio-temporal spectra $|\tilde{E}(k_{x},\lambda)|^{2}$ are plotted in Fig.~\ref{Fig:Spectral3-4-5-6-7}. Changing $\theta$ at fixed $\Delta\lambda$ results in a change of the spatial bandwidth $\Delta k_{x}$ and accordingly the transverse spatial width. The measured spectra are segments from the theoretically predicted conic sections: Fig.~\ref{Fig:Spectral3-4-5-6-7}(a,g) are ellipses, Fig.~\ref{Fig:Spectral3-4-5-6-7}(b) is a tangential line,  Fig.~\ref{Fig:Spectral3-4-5-6-7}(c-e) are hyperbolas, and Fig.~\ref{Fig:Spectral3-4-5-6-7}(f) is a parabola. They are all \textit{approximately} parabolas because of the limited $\Delta k_{x}$ utilized.

\begin{figure}[t!]
  \begin{center}
  \includegraphics[width=8.5cm]{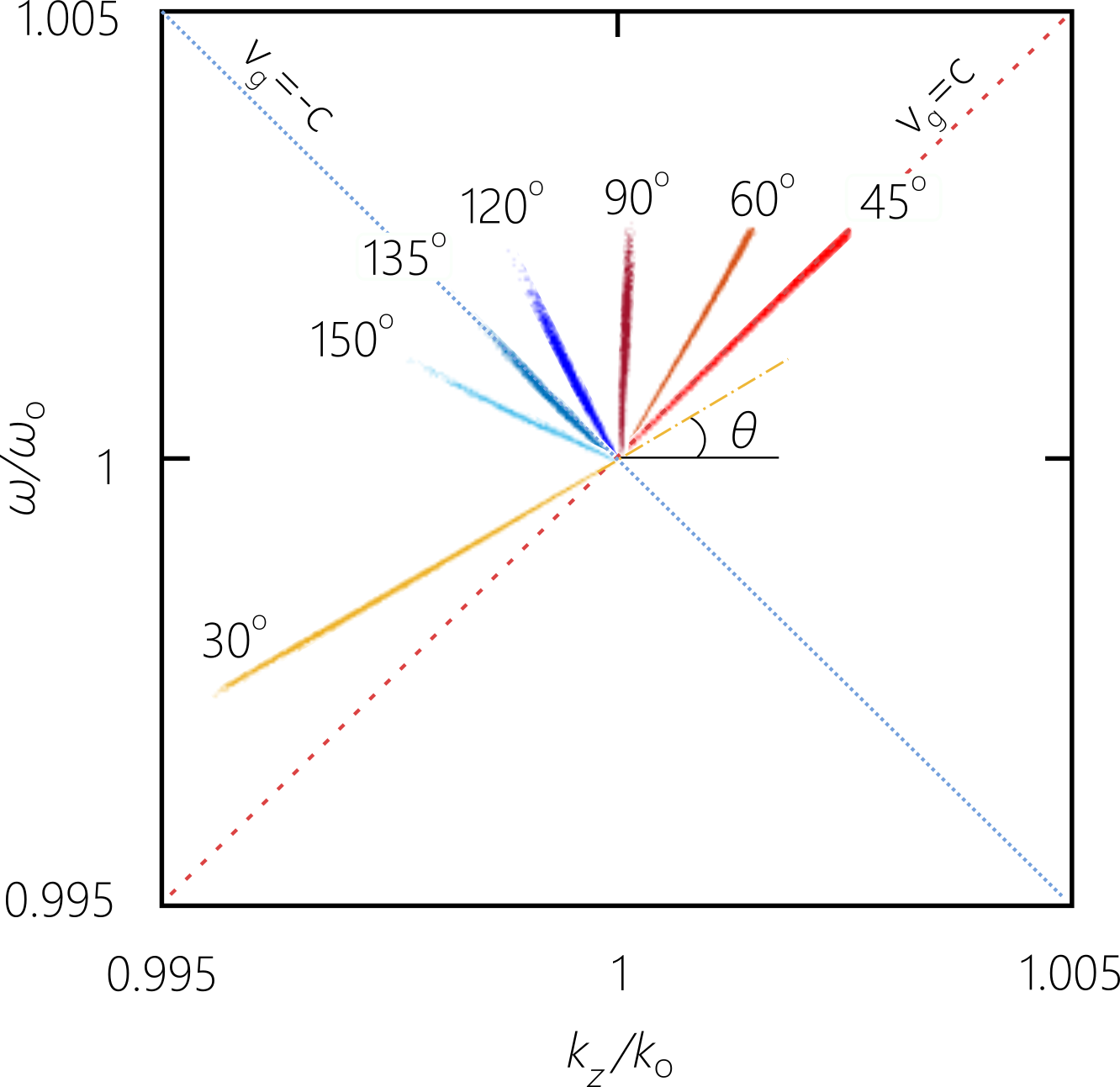}
  \end{center}
  \caption{Combined projections of the \textit{measured} spatio-temporal spectra for baseband ST wave packets with increasing $\theta$, Class-1 through Class-7, from Fig.~\ref{Fig:Spectral3-4-5-6-7}. The corresponding values of $v_{\mathrm{g}}$ with increasing $\theta$ (rotating counter-clockwise) are $0.577c$, $c$, $1.732c$, $\infty$ (formally; the value is limited by the spectral uncertainty), $-1.732c$, $-c$, and $-0.77c$; see \cite{Kondakci18unpub}.}
  \label{Fig:CombinedBaseband}
\end{figure}

Several observations are useful in interpreting the phase patterns $\Phi(x,y)$. First, as $\Delta k_{x}$ increases, higher spatial variations in $\Phi(x,y)$ are required; see lass-6 and Class-7 in Figs.~\ref{Fig:Spectral3-4-5-6-7}(f,g). Second, the \textit{sign} of the spectral curvature in the vicinity of $k_{x}\!=\!0$ determines the orientation of $\Phi(x,y)$. In Fig.~\ref{Fig:Spectral3-4-5-6-7}(a) for Class-1, the minimum wavelength is associated with $k_{x}\!=\!0$, whereas Class-3 through Class-7 have the opposite curvature compared to Class-1, and thus $k_{x}\!=\!0$ is located on the right (longer wavelengths). Third, because Class-2 wave packets are pulsed plane waves, $\Phi(x,y)$ is a constant [Fig.~\ref{Fig:Projections3-4-5-6-7}(b)]. Finally, in contrast to $\Phi(x,y)$ in Fig.~\ref{Fig:Spectral1-2}(a) utilized in synthesizing the monochromatic and separable pulsed beams, the positive and negative values of $k_{x}$ are arranged in the upper and lower halves of $\Phi(x,y)$, respectively. The spectral uncertainty $\delta\lambda$ is $\sim\!30$~pm, and we thus expect that the synthesized ST wave packets will be quasi-diffraction-free over an axial propagation distance of at least $\sim\!15$~mm \cite{Bhaduri18OE}.

\begin{figure}[t!]
  \begin{center}
  \includegraphics[width=8.5cm]{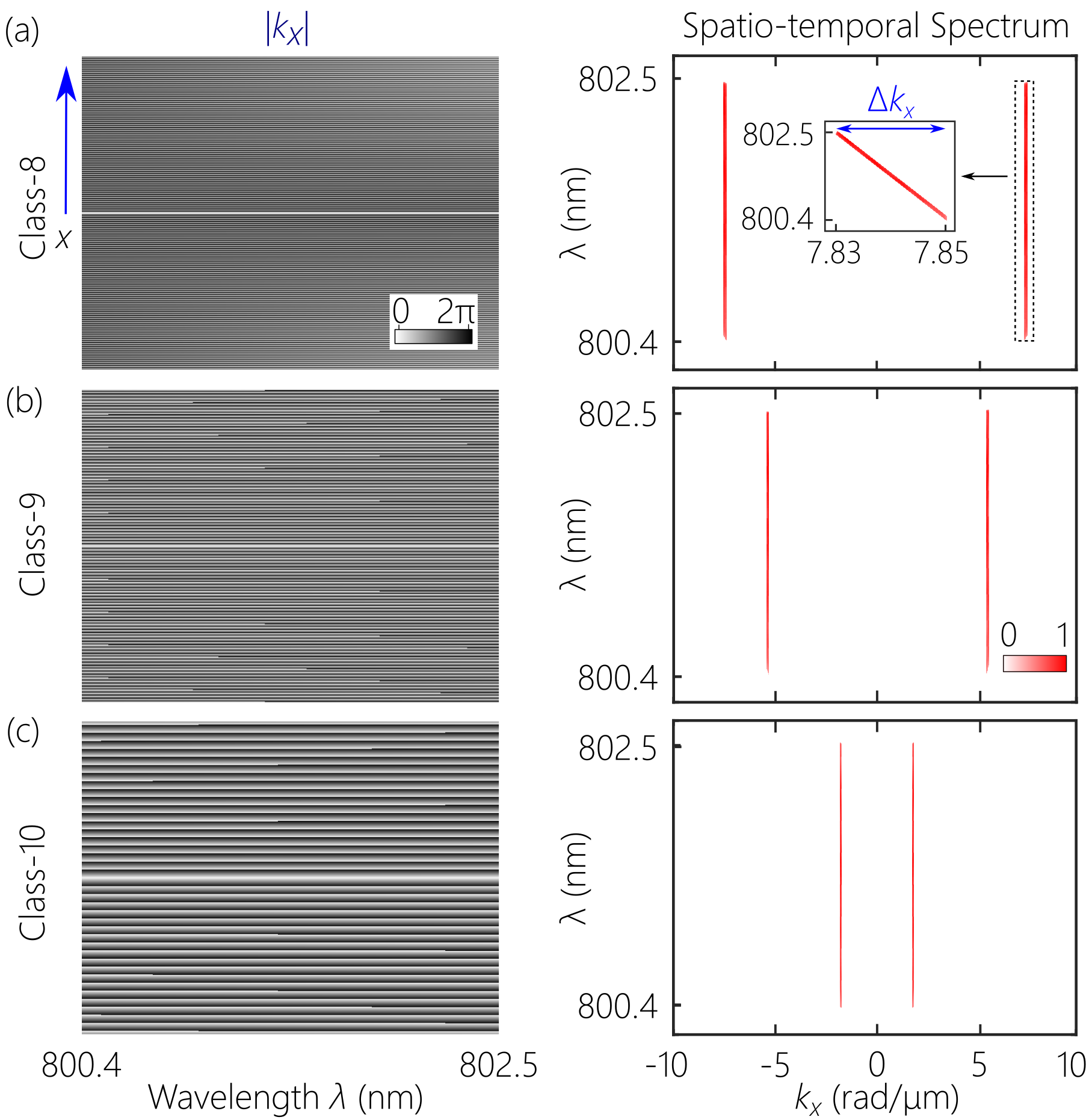}
  \end{center}
  \caption{Left and right columns show SLM phase distributions $\Phi(x,y)$ and the \textit{measured} spatio-temporal spectra $|\tilde{E}(k_{x},\lambda)|^{2}$ for sideband ST wave packets, respectively; $\Delta\lambda\!=\!2.1$~nm throughout. (a) Class-8 with $\theta\!=\!45^{\circ}$ and $\Delta k_{x}\!=\!0.02$~rad/$\mu$m; (b) Class-9 with $\theta\!=\!46^{\circ}$ and $\Delta k_{x}\!=\!0.01$~rad/$\mu$m; (c) Class-10 with $\theta\!=\!46^{\circ}$ and $\Delta k_{x}\!=\!0.01$~rad/$\mu$m. In all three classes, the spatial bandwidth $\Delta k_{x}$ is very small and not easily resolvable. The inset in (a) is one branch of the spatio-temporal spectrum to show the extent of the spatial bandwidth.}
  \label{Fig:Spectral8-9-10}
\end{figure}

We obtain from the spatio-temporal spectra of baseband ST wave packets projected onto the $(k_{x},\lambda)$-plane, as plotted in Fig.~\ref{Fig:Spectral3-4-5-6-7}, the corresponding spatio-temporal spectra projected onto the $(k_{z},\omega)$-plane, which we plot in Fig.~\ref{Fig:CombinedBaseband}. This plot highlights that the spectral projections onto the $(k_{z},\omega)$-plane are all straight lines -- despite their different appearance in the $(k_{x},\lambda)$-plane. Furthermore, Fig.~\ref{Fig:CombinedBaseband} confirms that the targeted spectral tilt angle $\theta$ for the representative of each class is correctly realized. The group velocity of each wave packet is $v_{\mathrm{g}}\!=\!c\tan{\theta}$, and the results plotted in Fig.~\ref{Fig:CombinedBaseband} span -- for the first time -- the whole gamut of subluminal, luminal, and superluminal values, both positive and negative. We have provided direct measurements of $v_{\mathrm{g}}$ in Ref.~\cite{Kondakci18unpub} using an interferometric method for a subset of this range. 

\subsection{Measurements of sideband ST wave packets}

We report in Fig.~\ref{Fig:Spectral8-9-10} the 2D phase distributions $\Phi(x,y)$ and the measured spatio-temporal spectra $|\widetilde{E}(k_{x},\lambda)|^{2}$ for \textit{sideband} ST wave packets, Class-8 through Class-10, at a fixed temporal bandwidth of $\Delta\lambda\!\sim\!2.1$~nm. The measured spectra are segments from the theoretically predicted conic sections: Fig.~\ref{Fig:Spectral8-9-10}(a) is a parabola, Fig.~\ref{Fig:Spectral8-9-10}(b) is a hyperbola, and Fig.~\ref{Fig:Spectral8-9-10}(c) a pair of lines. Because of the high $k_{x}$, however, the slopes of all the spatio-temporal spectra appear linear. Referring to the phase distributions $\Phi(x,y)$ in Fig.~\ref{Fig:Spectral8-9-10}, first note the absence of the constant phase of $k_{x}\!=\!0$ and of the slowly varying phases associated with low $k_{x}$, as expected in sideband ST wave packets in which the vicinity of $k_{x}\!=\!0$ is excluded. Second, the spatial variation in $\Phi(x,y)$ for sideband classes is much more rapid than for the baseband classes shown in Fig.~\ref{Fig:Spectral3-4-5-6-7} because of the large values of $k_{x}$ involved. Third, because sideband ST wave packets are restricted to the range $\tfrac{\pi}{4}\!<\!\theta\!<\!\tfrac{\pi}{2}$, the left-right orientation $\Phi(x,y)$ is always the same.

\begin{figure}[t!]
  \begin{center}
  \includegraphics[width=8.5cm]{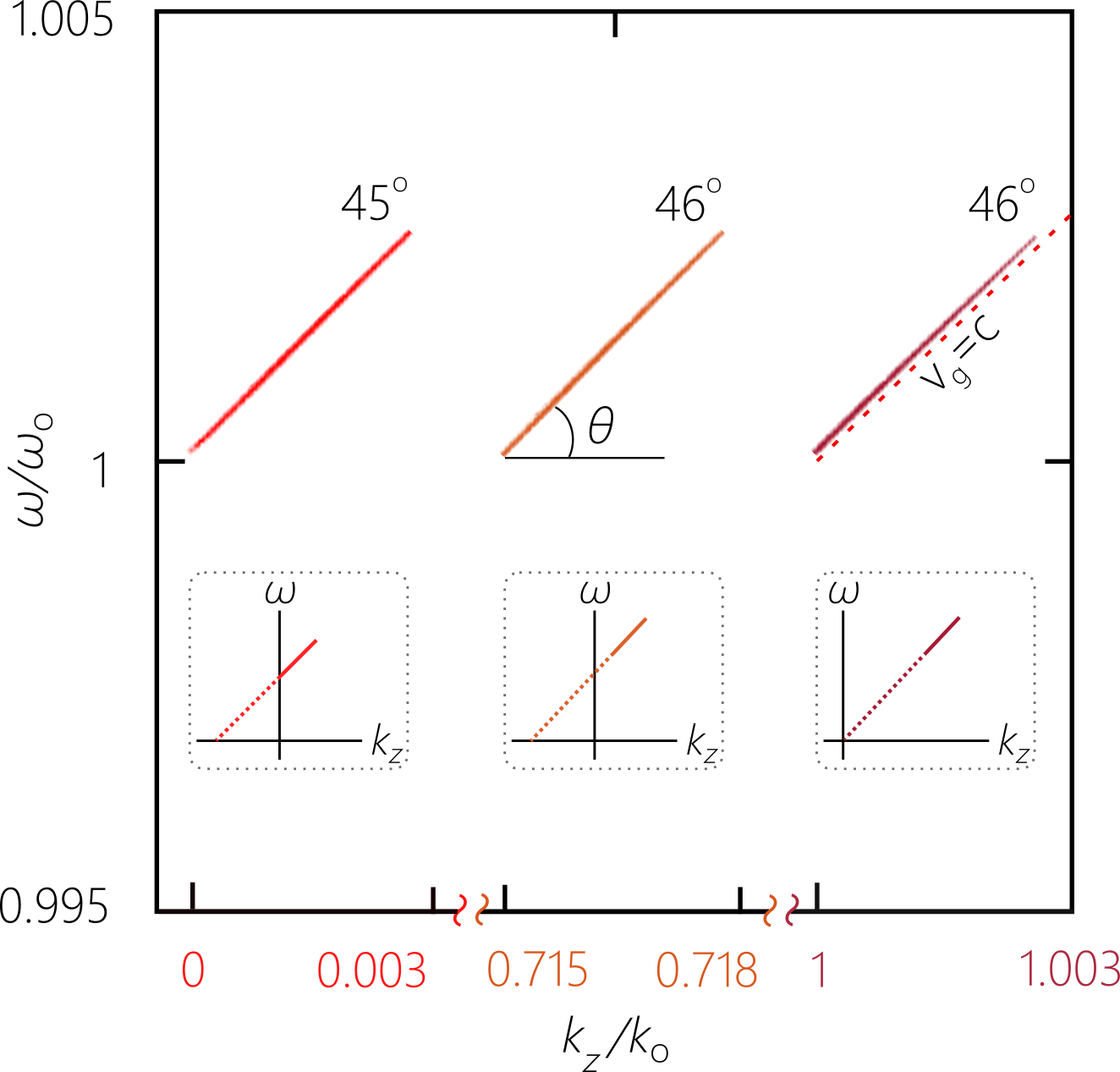}
  \end{center}
  \caption{Combined projections of the \textit{measured} spatio-temporal spectra for sideband ST wave packets, Class-8 through Class-10 from left to right, from Fig.~\ref{Fig:Spectral8-9-10}. The insets show the full $(k_{z},\omega)$ plane for each class. The deviation of $\theta$ from $45^{\circ}$, and thus the deviation of $v_{\mathrm{g}}$ from $c$ is minute because of the large values of $k_{x}$ involved in the spatio-temporal spectrum. The horizontal axis concatenates the very different ranges of $k_{z}$ associated with the three classes. The corresponding values of $v_{\mathrm{g}}$ are $c$, $1.036c$, and $1.036c$.}
  \label{Fig:CombinedSideband}
\end{figure}

We obtain from the spatio-temporal spectra of sideband ST wave packets projected onto the $(k_{x},\lambda)$-plane, as plotted in Fig.~\ref{Fig:Spectral8-9-10}, the corresponding spatio-temporal spectra projected onto the $(k_{z},\omega)$-plane, which we plot in Fig.~\ref{Fig:CombinedSideband}. Once again, the spectral projections onto the $(k_{z},\omega)$-plane are all straight lines. Crucially, Fig.~\ref{Fig:CombinedSideband} shows clearly the difficulty associated with tuning the spectral tilt angle $\theta$ of sideband ST wave packets because of the large values of $k_{x}$ involved. Therefore, it is expected that measured values of $v_{\mathrm{g}}$ of such wave packets will not deviate substantially from $c$. These results indicate that baseband ST wave packets are much more versatile with respect to the possibility of controlling their group velocity.

\subsection{Measurements of the axial propagation}

\begin{figure}[t!]
  \begin{center}
  \includegraphics[width=8.5cm]{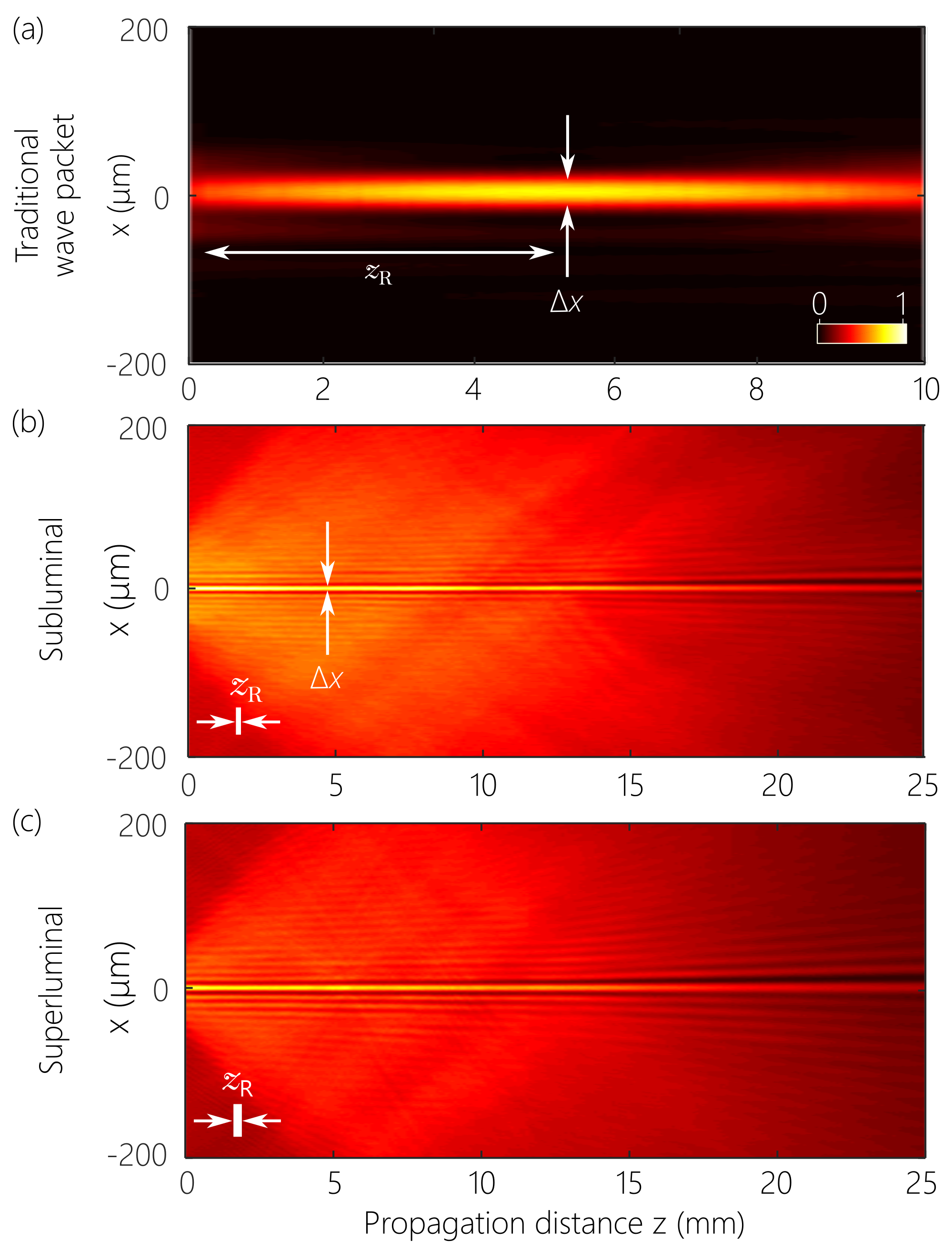}
  \end{center}
  \caption{Measured axial evolution of the time-averaged intensity $I(x,z)$ for (a) a separable pulsed beam having $\Delta\lambda\!=\!8.5$~nm, $\Delta k_{x}\!=\!0.09$~rad/$\mu$m, and $\Delta x\!=\!35$,~$\mu$m, corresponding to Fig.~\ref{Fig:Spectral1-2}(c); (b) a Class-1 ST wave packet with $\theta\!=\!30^{\circ}$, $\Delta\lambda\!=\!2.1$~nm, $\Delta k_{x}\!=\!0.48$~rad/$\mu$m, and $\Delta x\!=\!6$~$\mu$m, corresponding to Fig.~\ref{Fig:Spectral3-4-5-6-7}(a); and (c) a Class-3 ST wave packet with $\theta\!=\!60^{\circ}$, $\Delta\lambda\!=\!2.1$~nm, $\Delta k_{x}\!=\!0.36$~rad/$\mu$m, and $\Delta x\!=\!8$~$\mu$m, corresponding to Fig.~\ref{Fig:Spectral3-4-5-6-7}(c). The Rayleigh range $z_{\mathrm{R}}$ associated with the transverse width $\Delta x$ is identified graphically in each of the three cases, and are given by (a) 4.8~mm, (b) 150~$\mu$m, and (c) 250~$\mu$m. Note the different axial (horizontal) scale used in (a) with respect to (b) and (c).}
  \label{Fig:AxialPlots}
\end{figure}

Finally, we present measurements of the axial evolution of ST wave packets. We plot in Fig.~\ref{Fig:AxialPlots}(a) the time-averaged intensity distribution $I(x,z)$ for the \textit{separable} pulsed beam. As expected, diffractive spreading occurs away from the beam waist by the predicted Rayleigh range $z_{\mathrm{R}}\!=\!4.8$~mm (transverse spatial width $\Delta x\!=\!35$~$\mu$m). Similar results are observed for the monochromatic beam that has the same spatial spectrum as the separable pulsed beam. A wholly different behavior emerges for ST wave packets, as shown in Fig.~\ref{Fig:AxialPlots}(b) for a Class-1 ST wave packet at $\theta\!=\!30^{\circ}$ [Fig.~\ref{Fig:Spectral3-4-5-6-7}(a)], and in Fig.~\ref{Fig:AxialPlots}(c) that of a Class-3 ST wave packet at $\theta\!=\!60^{\circ}$ [Fig.~\ref{Fig:Spectral3-4-5-6-7}(c)]. We note first that the transverse beam profile is similar in both as a result of using the same incident femtosecond laser beam. Indeed, representatives for all the other baseband ST wave packets featured the same beam profile. Note that the larger spatial bandwidth $\Delta k_{x}$ employed with respect to the separable pulsed beam has produced a smaller beam width $\Delta x$, as expected, but a \textit{larger} propagation distance. Second, the sole observable difference is a change in the beam width $\Delta x$ which varies inversely with $\Delta k_{x}$ and depends on $\theta$, as confirmed in Figs.~\ref{Fig:AxialPlots}(b,c).

\section{Conclusions}

Our classification reveals some fundamental differences between the different classes of ST wave packets. For example, some classes are amenable to ultrawide temporal and spatial bandwidths (e.g., Class-3 through Class-5), while others have restrictions on the achievable bandwidths (e.g., Class-1, Class-6, and Class-7). Surprisingly, we find that the sign of the \textit{phase} velocity determines whether low spatial frequencies are allowable. As such, \textit{baseband} ST wave packets (Class-1 through Class-7) are associated with \textit{positive}-$v_{\mathrm{ph}}$, whereas \textit{sideband} ST wave packets (Class-8 and Class-9) are associated with \textit{negative}-$v_{\mathrm{ph}}$. Moreover, \textit{only} baseband ST wave packets (specifically, Class-5 through Class-7) can achieve negative group velocities. Finally, we have found that Class-1 ST wave packets allow a two-to-one association between $|k_{x}|$ and $\omega$ in some range of spatial frequencies, whereas all other classes maintain a one-to-one correspondence.

The conceptual and theoretical framework outlined here can be developed in a variety of directions. We have discussed ST wave packets that are $(2\!+\!1)$D in the form of a light sheet. An important question is whether the classification presented here extends to the $(3\!+\!1)$D case or whether new classes emerge. We conjecture that the current classification is in fact exhaustive since the transition from the $(2\!+\!1)$D case to the $(3\!+\!1)$D case involves exchanging the spatial frequency $|k_{x}|$ for the 2D transverse spatial frequency $k_{\mathrm{T}}\!=\!(k_{x}^{2}+k_{y}^{2})^{1/2}$. However, it is not clear at the moment how the experimental strategy described here can be extended to the synthesis of $(3\!+\!1)$D wave packets.

Recently, we have demonstrated that the tilting of spectral hyperplanes is associated with relativistic Lorentz transformations in frames moving relative to the source \cite{Kondakci18PRL}. Thus, transitioning from one class to another in our classification can be viewed as the result of a Lorentz boost; see also Refs.~\cite{Belanger86JOSAA,Saari04PRE,Longhi04OE,Wong17ACSP2}. In this sense, the ability to controllably synthesize ST wave packets opens up avenues for laboratory-scale studies of relativistic optical effects \cite{Bliokh12}. Furthermore, ST wave packets are a realization of `classical entanglement', which is the analog of multi-partite quantum entanglement applied to the correlations between the different degrees of freedom of a classical optical field \cite{Qian11OL,Kagalwala13NP,Abouraddy14OL,Aiello15NJP,Kagalwala15SR,Okoro17Optica}. Most studies of classical entanglement have focused on correlations between \textit{discretized} degrees of freedom, such as polarization and spatial modes, whereas ST wave packets are an example of classical entanglement between continuous degrees of freedom (spatial and temporal frequencies). Although we have focused here on coherent pulsed fields, there is nothing to prevent implementing all these classes of ST wave packets using broadband incoherent light instead \cite{Yessenov18Optica}. One may now consider the 10 classes studied here in the context of incoherent ST fields, whereupon the `speed of coherence' becomes tunable.

In conclusion, we have presented a comprehensive theoretical and experimental classification of all $(2\!+\!1)$D propagation-invariant ST wave packets in free space. In constructing this classification, we have made use of three criteria: the magnitude of the group velocity (subluminal, luminal, or superluminal); the sign of the group velocity (positive or negative); and whether the ST wave packet is `baseband' or `sideband' -- whether low spatial frequencies are allowed or forbidden, respectively. This classification reveals 10 distinct classes that we have described in detail and synthesized experimentally using an optical arrangement that combines spatial beam-modulation with ultrafast pulse-shaping. We have eschewed the traditional \textit{analytic} approach whereby specific solutions for the appropriate wave equation are sought, and have instead adopted a \textit{synthetic} strategy whereby an arbitrary spectrum is utilized in the wave packet's plane-wave expansion. This approach amounts to a realization of \textit{spatio-temporal Fourier optics} applied to the problem of synthesizing propagation-invariant wave packets.

\section*{Acknowledgments}
This work was supported by the U.S. Office of Naval Research (ONR) under contract N00014-17-1-2458.

\bibliography{diffraction}

\end{document}